\documentclass[reprint,
superscriptaddress,
 amsmath,amssymb,
 aps,
]{revtex4-2}

\usepackage{soul} 

\usepackage{graphicx}
\usepackage{dcolumn}
\usepackage{bm}
\usepackage{hyperref}
\usepackage{units}

\usepackage{physics}
\usepackage{color}
\usepackage{float}
\usepackage{placeins}

\definecolor{darkblue}{rgb}{0.031,0.282, 0.49}
\newcommand{\figlink}[1]{{\color[rgb]{0.031,0.282, 0.49}#1}}

\usepackage{mleftright} 

\hypersetup{
    colorlinks      = true,
    linkcolor       = darkblue,
    menucolor       = black,
    citecolor       = darkblue,
    urlcolor        = blue,
}

\renewcommand{\H}{\hat{H}}

\renewcommand{\a}{\hat{a}}
\newcommand{\ad}{\hat{a}^\dagger}

\newcommand{\bi}{\hat{b}_i}
\newcommand{\bid}{\hat{b}^\dagger_i}
\newcommand{\rhoh}{\hat{\rho}}

\newcommand{\ci}{\hat{c}_i}
\newcommand{\xc}{\hat{x}_c}

\newcommand{\xei}{\hat{x}_{e,i}}
\newcommand{\pc}{\hat{p}_c}

\newcommand{\pei}{\hat{p}_{e,i}}

\newcommand{\sm}{\hat{\sigma}_-}
\renewcommand{\sp}{\hat{\sigma}_+}

\newcommand{\smi}{\hat{\sigma}_{-i}}
\newcommand{\spi}{\hat{\sigma}_{+i}}

\newcommand{\wD}{\omega_d}         
\newcommand{\wc}{\omega_{c}} 
\newcommand{\we}{\omega_{e}}   
\newcommand{\wi}{\omega_i}
\newcommand{\Om}{\Omega}        

\newcommand{\mc}{m_c}
\newcommand{\me}{m_e}
\newcommand{\muc}{\Bar{\mu}_c}
\renewcommand{\qc}{q_c}
\newcommand{\gamc}{\gamma_c}
\newcommand{\game}{\gamma_e}

\newcommand{\gtwo}{g^{(2)}}
\newcommand{\nccl}{\langle n_c\rangle}
\newcommand{\ncbare}{\langle n^o_c\rangle}
\newcommand{\nenscl}{\langle n_{\rm ens}\rangle}
\newcommand{\ncweak}{\langle \hat{n}_c\rangle_{\rm weak}} 

\newcommand{\alphad}{\alpha_d}
\newcommand{\alphaens}{\alpha_{\text{ens}}}
\newcommand{\alphaeff}{\alpha_{\text{eff}}}
\newcommand{\gcol}{g_{\text{col}}}
\newcommand{\Omd}{\Omega_d}
\newcommand{\OmB}{\Omega_{\rm cr}}
\newcommand{\Omeff}{\Omega_{\text{eff}}}
\newcommand{\Omens}{\Omega_{\text{ens}}}

\newcommand{\figref}[1]{\mbox{Fig.~\ref{#1}}}

\newcommand{\secref}[1]{\mbox{Section~\ref{#1}}}

\newcommand{\appref}[1]{\mbox{Appendix~\ref{#1}}}
\renewcommand{\eqref}[1]{\mbox{Eq.~(\ref{#1})}}

\newcommand{\figpanel}[2]{Fig.~\hyperref[#1]{\ref*{#1}(#2)}}
\newcommand{\figpanels}[3]{Fig.~\hyperref[#1]{\ref*{#1}(#2)-(#3)}}
\newcommand{\figpanelNoPrefix}[2]{\hyperref[#1]{\ref*{#1}(#2)}}

\renewcommand{\exp}{\mathrm{e}}
\newcommand{\coop}{\text{C}}

\begin{document}

\preprint{APS/123-QED}
\title{Unconventional saturation effects at intermediate drive \\ in a lossy cavity coupled to few emitters}

\author{Therese Karmstrand}
\email{therese.karmstrand@chalmers.se}
\affiliation{Department of Microtechnology and Nanoscience (MC2), Chalmers University of Technology, 412 96 Gothenburg, Sweden}
\author{Benjamin Rousseaux}
\affiliation{Laboratoire Interdisciplinaire Carnot de Bourgogne, CNRS UMR 6303,
Universit\'e de Bourgogne, BP 47870, 21078 Dijon, France}

\author{Anton Frisk Kockum}
\affiliation{Department of Microtechnology and Nanoscience (MC2), Chalmers University of Technology, 412 96 Gothenburg, Sweden}

\author{Timur Shegai}
\affiliation{Department of Physics, Chalmers University of Technology, 412 96 Gothenburg, Sweden}

\author{G\"oran Johansson}
\affiliation{Department of Microtechnology and Nanoscience (MC2), Chalmers University of Technology, 412 96 Gothenburg, Sweden}

\date{\today}
\begin{abstract}
Recent technological advancements have enabled strong light-matter interaction in highly dissipative cavity-emitter systems. However, in these systems, which are well described by the Tavis--Cummings model, the considerable loss rates render the realization of many desirable nonlinear effects, such as saturation and photon blockade, problematic. Here we present another effect occurring within the Tavis--Cummings model: a nonlinear response of the cavity for resonant external driving of intermediate strength, which makes use of large cavity dissipation rates. In this regime, $(N+1)$-photon processes dominate when the cavity couples to $N$ emitters. We explore and characterize this effect in detail, and provide a picture of how the effect occurs due to destructive interference between the emitter ensemble and the external drive. We find that a central condition for the observed effect is large cooperativity, i.e., the product of the cavity and emitter decay rates is much smaller than the collective cavity-emitter interaction strength squared. Importantly, this condition does \textit{not} require strong coupling. We also find an analytical expression for the critical drive strength at which the effect appears. Our results have potential for quantum state engineering, e.g., photon filtering, and could be used for the characterization of cavity-emitter systems where the number of emitters is unknown. In particular, our results open the way for investigations of unique quantum-optics applications in a variety of platforms that neither require high-quality cavities nor strong coupling.
\end{abstract}
\maketitle

\section{Introduction}
At the heart of quantum optics lies the interaction of light with matter at the level of individual quanta.
As a result of the light-matter interaction between a single or an ensemble of two-level emitters and a resonant single-mode cavity, the emitters introduce nonlinearity to the otherwise linear cavity spectrum. This nonlinearity results in a splitting of eigenenergies known as the Jaynes-- and Tavis--Cummings ladders~\cite{Jaynes_Cummings_1963, tavis_cummings_1968}.
Under weak cavity and emitter excitation, one effect of these ladders is vacuum Rabi splitting in the spectrum of the system. Three other well-known quantum-optical effects also arise from this nonlinearity: saturation~\cite{Sanchez_1983}, photon blockade~\cite{Imamoglu_1997}, and unconventional photon blockade~\cite{Leonski_2004, Liew_2010,Flayac_2017}. These effects are all of great interest for quantum control of light fields with important applications such as single-photon switches~\cite{Volz2012, chen_all-optical_2013, shomroni_all-optical_2014, sun_single-photon_2018, munoz-matutano_all_2020} and transistors~\cite{Chang_2007,hwang_single-molecule_2009,chen_all-optical_2013,sun_single-photon_2018} and the generation of specific quantum states~\cite{law_arbitrary_1996,plenio_cavity-loss-induced_1999,Kim_1999,Pelton_2002, ritter_elementary_2012, Strauch2012, Muller_2015,You_2020}. 
In this paper, we demonstrate yet another effect, reminiscent of the saturation effect, which shows potential for applications in, e.g., quantum state engineering or the characterization of the number of quantum emitters in the cavity.

The saturation effect occurs when an emitter or nonlinear medium in a cavity cannot absorb more photons and thus has become saturated. In the spectrum, this is revealed as a merging of the vacuum Rabi doublet into a single Lorentzian peak at the cavity resonance when increasing the intracavity field~\cite{Sanchez_1983}. Ideally, a single photon incident on the system is needed to saturate a single emitter in the cavity. 
In this case, single-photon saturation could implement, e.g., a single-photon transistor~\cite{Chang_2007} or a single-photon sensor~\cite{varnava_how_2008,hadfield_single-photon_2009}.
Reference~\cite{Pscherer_2021} demonstrates experimental progress approaching the single-photon limit. However, the saturation effect is typically associated with a very strong drive.  
The need for a strong drive is due to the generally low probability for photon-emitter interaction~\cite{Chang_2014} and, in the many-emitters case, to the fact that the entire medium must be saturated~\cite{Gripp_2010}.
This is problematic for applications, especially if the systems exhibit large dissipation rates.  

In photon blockade~\cite{Imamoglu_1997}, on the other hand, the anharmonicity in the spectrum blocks the absorption of a subsequent photon.
The effect occurs for a resonant drive on one of the polariton transitions.
A characteristic of photon blockade is nonclassical photon-counting statistics, which can be probed via the normalized second-order correlation function $\gtwo$ in the weak drive regime~\cite{Hanbury_1956, Loudon_2000, Miranowicz_2010}. Two signatures of nonclassical light are photon antibunching [$\gtwo(\tau)>\gtwo(0)$] and sub-Poissonian photon statistics [$\gtwo(0)<1$]~\cite{Zou_1990,Vogel_Welsch_2006}. Thus, photon blockade could be exploited for the generation of nonclassical photon states, e.g., a single-photon source. The single-photon blockade has been extensively explored theoretically~\cite{Tian_1992,Imamoglu_1997,Brecha_1999,Werner_1999,Rabl_2011,Carmichael_2015} as well as demonstrated experimentally~\cite{Kim_1999,Birnbaum_2005, Faron_2008}. 
Stimulated by the potential for quantum state engineering including more than one photon~\cite{Chang_2014}, there have recently also been several works on the multi-photon blockade~\cite{Shamailov_2010,Miranowicz_2013,Radulaski_2017,Hamsen_2017, Zou_2020}.  Additionally, a break-down of the photon blockade has been observed for strong external driving\cite{Carmichael_2015,Alsing_1991,Alsing_1992}, and has been studied with mean field theory in the limit of large quantum emitter numbers \cite{gutierrez-jauregui_dissipative_2018}. 
One basic condition for both single- and multi-photon blockade is that the decay rates of the system should be much smaller than the cavity-emitter interaction strength. These conditions require high-quality cavities as well as small emitter dephasing. For that reason, demonstration of photon blockade in dissipative systems remains difficult.

An alternative approach to the generation of nonclassical states of light, exploiting the anharmonic Jaynes-- or Tavis--Cummings spectrum, is through the so-called unconventional photon blockade~\cite{Leonski_2004, Liew_2010,Flayac_2017}. In contrast to traditional photon blockade, the unconventional photon blockade effect relies on the interference between two transition pathways (see, e.g., Refs.~\cite{Bamba_2011,Majumbar_2012,You_2020}) when the drive is tuned in between the two polariton transitions. Thus, being an interference effect, the overlap due to broader transition linewidths can be exploited.
Similar to photon blockade, unconventional photon blockade displays nonclassical photon statistics in $\gtwo$ measurements in the weak drive regime. Originally, unconventional photon blockade was found for coupled Kerr resonators~\cite{Leonski_2004, Miranowicz_2006}. Since then, it has been
predicted~\cite{Bamba_2011, You_2020, Ridolfo_2010} and
demonstrated experimentally~\cite{Radulaski_2017,Snijders_2018} with dissipative cavity-emitter systems described by the driven Jaynes-- and Tavis--Cummings Hamiltonians.
It has also been predicted for large ensembles of emitters provided large enough individual cavity-emitter interaction strength~\cite{Saez-Blazquez_2017,Saez-Blazquez_2018}.
Nevertheless, demonstrating unconventional photon blockade remains difficult, due to fast oscillations of $\gtwo$ that exceed the resolution of state-of-the-art detectors and the requirement of fine-tuning of intrinsic system parameters~\cite{Flayac_2017}.

In this work, we demonstrate a different approach to harness the nonlinearity introduced by one or a few two-level emitters interacting with a dissipative cavity. Our scheme is simple, employing a continuous-wave (CW) coherent drive, requiring only detection of the steady-state cavity population. We base our analysis on numerical solutions of the corresponding Lindblad master equation. We use the rotating-wave approximation for the drive and coupling terms, but otherwise no further approximations that would limit us to the weak drive regime~\cite{Bamba_2011,Saez-Blazquez_2017,Saez-Blazquez_2018}. Therefore, we can explore the intermediate drive regime, where we find a saturation-like effect on the cavity population, due to destructive interference between two excitation pathways. The cavity can be excited either directly by the drive or by the excited emitters. The interference between these two transition pathways has similarities with the interference that gives rise to unconventional photon blockade. Therefore, we name the effect observed here \textit{unconventional saturation}.

The unconventional saturation effect is revealed in the cavity response to resonant driving of intermediate strength and arises due to the intermittent saturation of the destructive interference, leading to direct cavity excitation. Already visible in the weak-excitation regime, well before traditional saturation, the effect leads to a strong nonlinear dependence of the intracavity field on the drive strength. Moreover, it is \textit{not} limited to strong cavity-emitter coupling. Instead, we find that the basic requirements for observing unconventional saturation are: few quantum emitters, large cooperativity $C\equiv 4 \gcol^2 /\gamc \game$, and intermediate drive strengths. The second condition, large $C$, is naturally found in many lossy cavities where the cavity decay rate $\gamc$ is large compared to the emitter decay rate $\game$, such that they fulfill $\gamc \game \ll 4 \gcol^2$. Here, $\gcol$ is the collective cavity-emitter interaction strength.  In comparison to unconventional photon blockade, which often involves systems with small $C \approx 0.5 {\rm -} 2$ and weak driving, the unconventional saturation effect becomes notable for $C \gtrsim 10$ with intermediate drive strengths and grows more prominent for higher $C$. 

The signature of unconventional saturation is the dominance of $(N+1)$-photon processes in scattering 
from an $N$-emitter-cavity system. 
Somewhat hand-wavingly, the emitter ensemble can be seen as a saturable mirror, which only can reflect states with up to $N$ photons. We identify the origin of this effect as the same type of quantum interference that explains unconventional photon blockade.
Nevertheless, the fact that unconventional saturation can be detected in steady-state scattering could facilitate a more straightforward experimental demonstration than the more elaborate photon-correlation measurement $\gtwo$ needed for verifying unconventional photon blockade. Moreover, as opposed to vacuum Rabi splitting and photon-blockade techniques, our approach unambiguously differentiates between different numbers of emitters with the same collective interaction strength.
This property makes it a promising scheme for characterization of cavity-emitter systems where the number of emitters is unknown, e.g., counting of NV-centers in diamond~\cite{Doherty_2013,Chang_2016}, localized emitters in hBN~\cite{Vogl_2019,Fournier_2021}, or molecules in a Fabry-P\'erot cavity~\cite{Mony2021}, as well as for the verification of fundamental differences between single- and few-quantum-emitter systems. Other possible applications are within technologies such as quantum imaging~\cite{Moreau_2019}, quantum metrology~\cite{Giovannetti_2011} and more, which rely on the generation of nonclassical light fields.

Besides being a --~to our knowledge~-- novel quantum-optical effect, we also see the potential for the use of unconventional saturation for progressive quantum state engineering that could find a natural place in hybrid quantum systems, similar to the setup proposed in Ref.~\cite{Chang_2007}. The generation of specific quantum states of light with dissipative systems has already been proposed for other setups, including single-~\cite{Lindkvist_2014, Dhar_2018} and multi-photon~\cite{Eleuch_2012, Muller_2015, You_2020} generation. We believe that our work offers a foundation for further explorations of hitherto unknown effects that could complement and improve existing schemes. Our results already suggest a form of photon filtering that can be achieved using a setup (CW drive and scattering) that is simpler compared to many other schemes. 

A potential platform for demonstrating unconventional saturation is hybrid light-matter systems using, e.g., broad-linewidth surface plasmons and narrow-linewidth localized two-level emitters. 
Light-matter hybridization is a growing research field that utilizes hybridized states of light and matter such as surface plasmons as the carrier of the photonic component.
Part of the attraction of such setups is the sub-wavelength confinement of the light mode that can greatly enhance the interaction with optical emitters~\cite{Chang_2006, Baranov_2018, Bisht_2019, Stuhrenberg_2018}. Even ultrastrong coupling~\cite{Kockum2019,kuisma2022ultrastrong} has been demonstrated~\cite{Baranov_2020} in these systems. 
The potential of hybrid systems for quantum technology has already been demonstrated in Ref.~\cite{Chang_2007}, proposing a single-photon optical transistor. Lastly, the fact that strong coupling between cavities and emitter ensembles can be observed at room-temperature in these dissipative hybrid systems, motivates the search for observation and possible application of quantum-optical phenomena beyond cryogenic temperatures~\cite{rousseaux2018quantum,Palstra_2019,zasedatelev_single-photon_2021,zasedatelev_room-temperature_2019, heintz_few-molecule_2021}.

This article is organized as follows. Our theoretical framework is presented in \secref{sec:theory}, including the driven Tavis--Cummings model (\secref{sec:TCmodel}), the master equation used for numerical calculations (\secref{sec:ME}), scattering from the cavity (\secref{sec:Scattering}), 
and an analog classical coupled-oscillator model used for analytical calculations in the weak-drive regime (\secref{sec:COmodel}). In \secref{sec:results}, we present the results from our explorations of the driven Tavis--Cummings model, which show a saturation-like response of the cavity population in the intermediate drive regime. First, we study the spectrum in \secref{sec:Spectrum} and note that there is a sweet spot for quantum effects on resonance, which differentiates between different system sizes. Thereafter, the system response is examined for resonant driving in the weak- to intermediate- drive regimes in \secref{sec:MeanCavityResponse}. The observed nonlinear response is analyzed in \secref{sec:EffectiveDrive} in terms of an effective drive acting on the cavity. In \secref{sec:CriticalDrive}, we build on this description to derive an analytical expression for the critical drive strength required for entering the nonlinear regime. We also find a figure of merit for the observed effect in \secref{sec:FOM} and show that it can be explained by quantum interference effects in \secref{sec:QuantumInterference}. Finally, the conclusions from our investigations are presented in \secref{sec:Conclusions}. We give additional details for some calculations in appendices: Appendix~\ref{sec:app_COmodel} shows the mapping between the classical and quantum models used in \secref{sec:COmodel}, Appendix~\ref{app:Laser} reviews the quantum theory for a propagating laser beam, and Appendix~\ref{app:add_sim} contains plots further demonstrating the effect of coupling strengths and cooperativity on the unconventional saturation.

%
%
\section{Theoretical framework}
\label{sec:theory}

\subsection{Coherently driven Tavis-Cummings model}
\label{sec:TCmodel}

\begin{figure}
    \includegraphics[width=\linewidth]{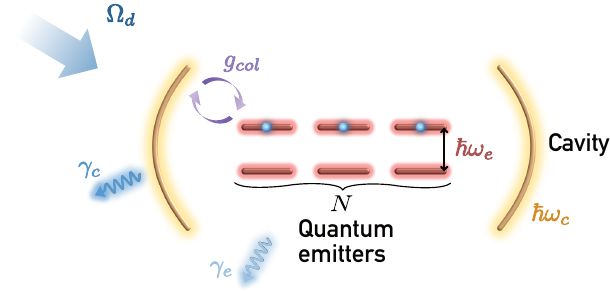}
    \caption{Schematic illustration of the open Tavis-Cummings system under investigation.}
    \label{fig:TCHamiltonian}
\end{figure}

Figure~\ref{fig:TCHamiltonian} shows a schematic illustration of the driven dissipative Tavis--Cummings system considered in this work. The Tavis--Cummings model describes the dynamics of an ensemble of $N$ identical quantum emitters interacting with a common single-mode cavity field~\cite{tavis_cummings_1968}. No interaction between the individual emitters is included, which is motivated in circumstances where the cavity-emitter interaction is the dominant interaction governing the dynamics. Including a coherent drive $\bar{\mathcal{E}}\cos(\wD t)$, with spatial amplitude $\bar{\mathcal{E}}$ and frequency $\wD$, the driven Tavis--Cummings Hamiltonian can be written within the rotating-wave approximation (RWA) as
\begin{eqnarray}
    \H_{TC} = && \hbar \wc \ad \a + \frac{\hbar \Omd}{2}\left(\ad e^{-i\wD t} + \a e^{i\wD t}\right)\nonumber \\ 
    && + \sum_{i=1}^N \left[ \hbar\we\spi\smi + \hbar g(\ad \spi + \a \smi)\right]. 
    \label{eq:Htc}
\end{eqnarray}
Here, $\a$ and $\ad$ are annihilation and creation operators, respectively, for the cavity mode, $\wc$ is the cavity frequency, $ \smi$ and $\spi$ are the Pauli lowering and raising operators, respectively, for the $i$th quantum emitter, $\we$ is the transition frequency of the emitters, $\Omd$ is the strength of the cavity drive, and $g$ is the strength of the coupling between the cavity mode and a single quantum emitter.

The cavity drive strength is given by $\Omd= (\bar{a}\cdot \bar{\mathcal{E})}/\hbar$, where the spatially dependent parameter $\bar{a}$ is cavity-specific. Thus, the exact form of $\Om_d$ is determined by the explicit drive and cavity configuration.
No external driving of the emitters is considered. This assumption is natural for most experimental setups where the emitters are located inside the cavity, but works as well for open cavities such as plasmonic nano-cavities that typically have much larger transition dipole moments than most quantum emitters.
Furthermore, spatial variations of the cavity-emitter dipole interaction is neglected. Thus, we take $g=\bar{\mu}_e\cdot \bar{\mathcal{E}_c}/\hbar$ for all emitters, with the transition dipole moment $\bar{\mu}_e$ interacting with the cavity field with amplitude $\bar{\mathcal{E}}_c$. This approximation is sufficient for many situations involving only a few localized quantum emitters, and in situations where the emitters are small compared to the cavity. With equal interaction rates $g$, the structure of the interaction term in the Tavis--Cummings Hamiltonian [Eq.~(\ref{eq:Htc})] leads to the collective interaction strength $g_{col} = \sqrt{N}g$ between the cavity and the collective bright mode of the emitter ensemble.

\subsection{Master equation}
\label{sec:ME}
In this work, the scattering from the system under weak to intermediate driving 
is investigated. To solve for the cavity-emitter state including dissipation, an open-quantum-system approach is employed, using the master equation
\begin{equation}
        \dot{\rhoh} = -\frac{i}{\hbar} \mleft[\H_{TC},\rhoh\mright] + \gamc \mathcal{D}_{\a}\mleft[\rhoh\mright] + \sum_{i=1}^N \game \mathcal{D}_{\smi}\mleft[\rhoh\mright].
        \label{eq:ME}
\end{equation}
Here the operator $\mathcal{D}_{\hat{o}}[\cdot] = \hat{o}\cdot \hat{o}^\dagger -\frac{1}{2}\acomm{\hat{o}^\dagger \hat{o}}{\cdot}$ acting on the density matrix $\rhoh$ is the standard Lindblad superoperator for dissipation associated with the operator $\hat{o}$~\cite{Lindblad1976}. With this master-equation approach, it is also possible to treat the case of strong driving, for which the traditional saturation effect would be found.

The first term in Eq.~(\ref{eq:ME}) describes coherent evolution with the Tavis--Cummings Hamiltonian. 
The second term describes radiative and non-radiative dissipation of the cavity mode, making the total dissipation rate $\gamc=\gamma^{r}_{c}+\gamma^{nr}_{c}$. In the third term, the individual dissipation rates $\gamma_e$ for the emitters are assumed to be equal. The form of Eq.~(\ref{eq:ME}) neglects the contribution of thermal photons to the system dynamics and is therefore valid for low temperatures or high-frequency quantum systems with $\hbar\omega_c, \hbar\omega_e \gg k_B T$, such that thermal fluctuations do not particularly affect the dynamics. In experimental realisations, this condition is naturally met, e.g, for optical frequencies at room temperature. 

A more compact way of writing Eq.~(\ref{eq:ME}) is in terms of the Liouvillian superoperator:
\begin{equation}
    \mathcal{L}[\cdot]=-\frac{i}{\hbar} \left[\H_{TC},\cdot\right] + \gamc \mathcal{D}_{\a}\left[\cdot\right] + \sum_{i=1}^N \game \mathcal{D}_{\smi}\left[\cdot\right].
\end{equation} 
Then, the task of finding the steady state is reduced to the eigenvalue problem
\begin{equation}
     \mathcal{L}\left[\rhoh_{ss}\right] = 0
     \label{eq:ME_ss}
\end{equation}
with a Hermitian density operator $\hat{\rho}_{ss}$ satisfying the normalisation condition 
\begin{equation}
    \Tr{\hat{\rho}_{ss}} = 1.
\end{equation}

\subsection{Probing the cavity}
\label{sec:Scattering}
For applications in quantum photonics, the scattering from the system is of great interest. In cavity-emitter systems where the cavity interacts much more strongly with the environment, the collection of emitted photons from the emitters may be neglected. This complies with the condition $\gamma_c^r \gg \game$, which is what is considered in this work. Moreover, in most experimental setups, the collection of emitted photons from the driven system can be located such that the incident laser field is filtered out. The collected scattering $S$ from the system will therefore be proportional to the radiative cavity decay rate and the average cavity population:
\begin{equation}
    S \propto \gamc^r \expval{\ad \a}.
    \label{eq:Scattering}
\end{equation}

\subsection{Analogue classical coupled oscillator model in weak drive regime}
\label{sec:COmodel}
For adequately weak drive, much of the phenomenology associated with coupled cavity-emitter systems can be described by a classical coupled-oscillator (CO) model~\cite{Novotny_2010,Torma_2015,Alzar_1995}. Here a CO model will be used for comparison when analyzing the quantum effects that arise beyond the weak drive regime. 
  
The CO model considered involves $N+1$ mechanically coupled masses on springs. The corresponding classical coupling constant and drive strength are $-2g\sqrt{m_c m_e \wc \we}$ and $\Omd\sqrt{2m_c\hbar\wc}$, respectively. For simplicity, the cavity and emitter masses, $m_c$ and $m_e$, are set to 1. The mapping of the quantum parameters to the classical model can be found Appendix~\ref{sec:app_COmodel}.
Letting index $0$ denote the oscillator representing the cavity mode and index $1, ..., N$ the emitters, the equations of motions for the classical analog of $N$ identical emitters coupled to a  coherently driven cavity mode are
\begin{align}
    \label{eq:x0} 
    &\ddot{x}_0 + \gamc \dot{x}_0 + \wc^2 x_0 + \sum_{i = 1}^N 2g\sqrt{\wc\we} x_i = \Omd\sqrt{2\hbar \wc}\cos{(\wD t)} , \\
    \label{eq:xi}
    &\ddot{x}_i + \game \dot{x}_i + \we^2 x_i + 2g\sqrt{\wc\we} x_0 = 0, \quad  i = 1,\, 2,\, ...,\, N .
\end{align}

The set of coupled equations~(\ref{eq:x0})-(\ref{eq:xi}) is easily solved by making the ansatz $z_i=C_ie^{i\wD t}$ for all $i=0,.., N$ and noting that $x_i = \Re{z_i}$ and $\cos{(\wD t)} = \Re{e^{i\wD t}}$. The solutions for the amplitudes are
\begin{align}
    \label{eq:C0}
    &C_0 =\frac{\Omd\sqrt{2\hbar\wc}(\we^2-\wD^2+i\wD\game)}{(\wc^2-\wD^2+i\wD\gamc)(\we^2-\wD^2+i\wD\game)-4Ng^2\wc\we} ,\\
    \label{eq:Ci}
    &C_i =\frac{-2g\sqrt{\wc\omega_e}\Omd\sqrt{2\hbar\wc}}{(\wc^2-\wD^2+i\wD\gamc)(\we^2-\wD^2+i\wD\game)- 4Ng^2\wc\we}.
\end{align}

Equations~(\ref{eq:C0}) and (\ref{eq:Ci}) can be used to calculate the classical oscillator energies   $E_{c/e} = \frac{1}{2} \omega^2_{c/e}|C_{0/i}|^2$, which can be compared with the average energies $E^{qm}_{c} = \hbar \wc \langle \ad \a \rangle$ and $E^{qm}_{e,i} = \hbar \we \langle \spi \smi \rangle$ in the cavity mode and emitter ensemble, respectively, calculated using the Tavis--Cummings model. Since the average energies in both models must be the same, the classical analogue to the populations is given by
\begin{align}
    \label{eq:npl}
    & \nccl = \frac{\wc^2|C_0|^2}{2 \hbar \wc}, \\
    \label{eq:nqe}
    & \nenscl = \sum_i^N \frac{\we^2|C_i|^2}{2\hbar \we} = \frac{N\we^2|C_i|^2}{2\hbar \we}.
\end{align}
Equations~(\ref{eq:npl}) and (\ref{eq:nqe}) will be useful for comparing the classical and quantum results in this article. Note that Eq.~(\ref{eq:nqe}) represents the total ensemble average population.

%
%
\section{Unconventional saturation effect at resonant driving}
\label{sec:QuantumNonLinearResp}
\label{sec:results}

\subsection{Scattering spectrum}
\label{sec:Spectrum}

\begin{figure}
    \includegraphics[width=\linewidth]{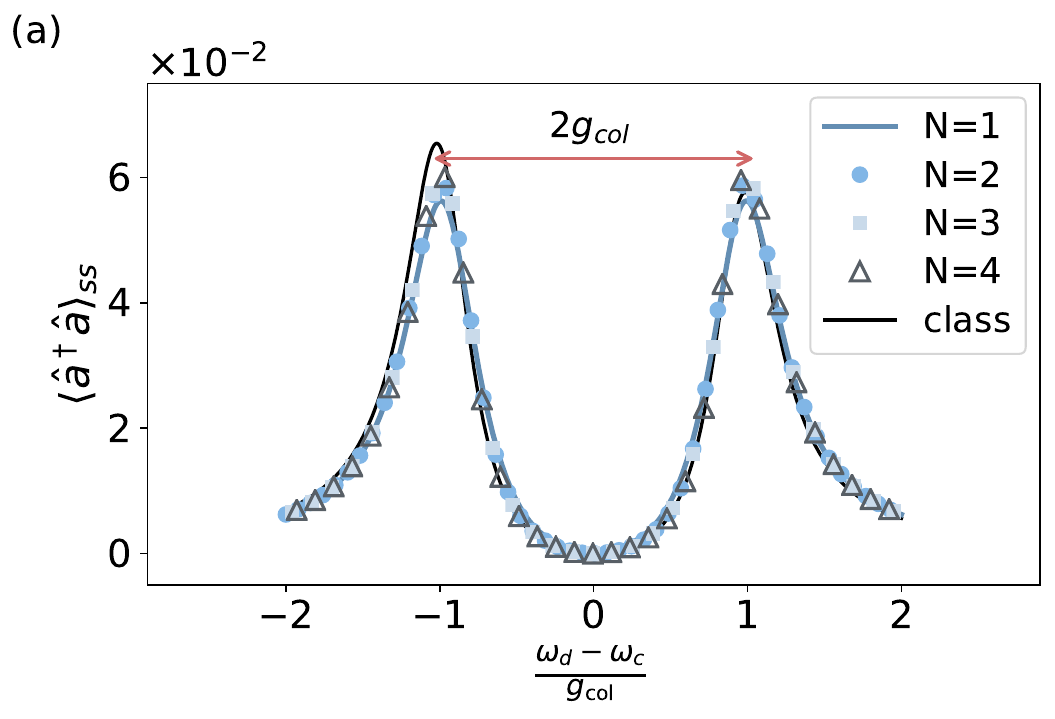}\\
    \includegraphics[width=\linewidth]{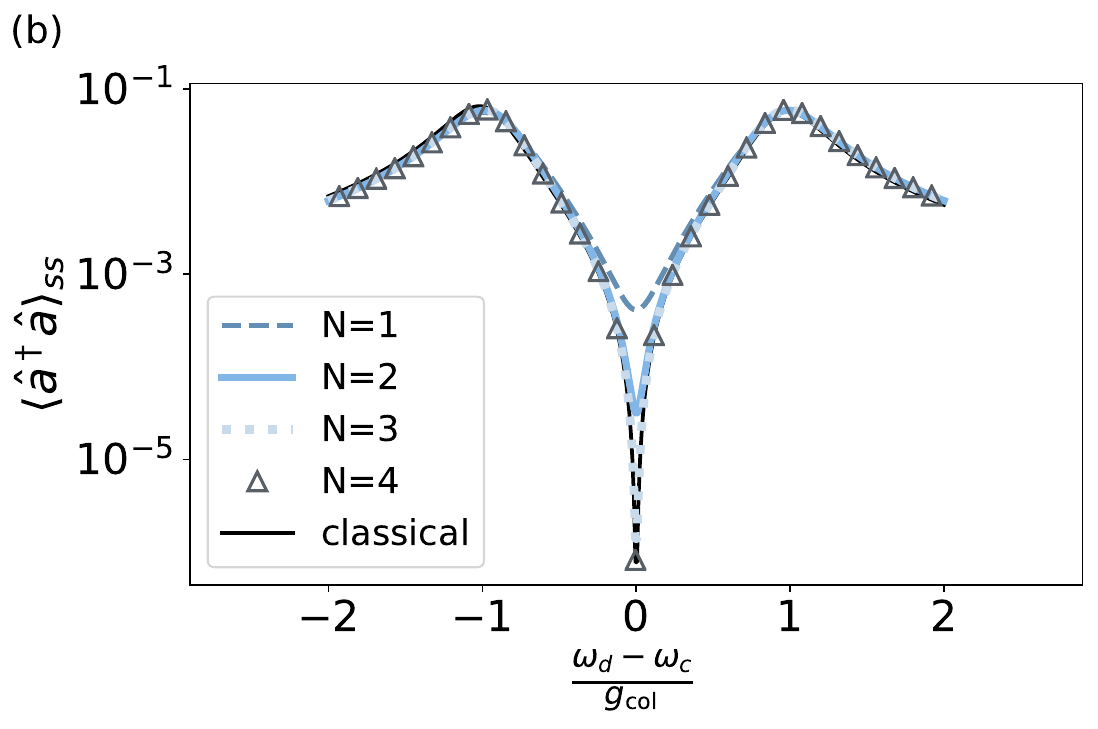}%
    \caption{Master-equation calculations of the cavity populations in the steady state for coupled cavity-$N$-emitter systems with the same collective interaction strength $\gcol$. The spectra are compared with the analogue, classical calculation.
    (a) The spectra for $N = 1-4$ quantum emitters show only minor differences between different $N$ and the classical CO model when plotted on a linear scale.
    (b) The spectra viewed on a logarithmic scale, on the contrary, show considerable differences of several orders of magnitude for resonant driving.}
    \label{fig:spectrum}
\end{figure}

Large loss rates generally limit experimental investigations to weak excitation, $\langle \ad \a \rangle \ll 1$. In this regime, strong coupling with the emitter ensemble will lead to vacuum Rabi splitting in the spectrum. This effect can be seen in Fig.~\ref{fig:spectrum}\figlink{(a)} for the steady-state cavity population $\langle \ad \a \rangle_{ss} = \Tr{\ad \a \hat{\rho}_{ss}}$ under continuous driving within the weak-excitation regime, with emitters on resonance with the cavity ($\wc=\we$). The steady state $\hat{\rho}_{ss}$ is found from the master equation by solving Eq.~(\ref{eq:ME_ss}) for $N=1,2,3$, and $4$ quantum emitters, and is compared with the classical solution $\langle n_{c}\rangle$ given by Eq.~(\ref{eq:npl}). 
As can be seen, there are only small differences in the spectra between different $N$ and the classical solution.   

On the other hand, examining the same spectra on a logarithmic scale in Fig.~\ref{fig:spectrum}\figlink{(b)}, large deviations (several orders of magnitude) from the classical model can be seen when the drive is resonant with the cavity and the emitters.
Despite having the same collective interaction strength $\gcol$, large differences can also be seen between the spectra for different numbers of emitters in the ensemble. 
The spectrum for $N=1$ shows the largest deviation from the classical case; adding more emitters yields spectra approaching the classical response.
Thus, we have found a sweet spot for quantitative quantum effects that differentiate between different emitter-ensemble sizes $N$ in the weak-excitation regime. In fact, it turns out that a strongly $N$-dependent nonlinear response can be accessed for resonant driving in the steady state, as will be shown  in the next section.  

For this simulation and throughout the main text, the emitters are taken to be on resonance with the cavity mode, i.e., $\omega_{e}=\omega_{c}$. The other parameters used in Fig.~\ref{fig:spectrum} were $\gamma_c/\omega_c=0.03$, $\gamma_e/\omega_c=0.0003$, $\gcol/\omega_c=0.03$, and $\Omd/\gcol=0.25$.

\subsection{Mean cavity response for increasing drive rate}
\label{sec:MeanCavityResponse}
Encouraged by the visible quantum effects on resonance in the spectrum, we here further explore the optical response of the Tavis--Cummings model for resonant driving. The spectra in Fig.~\ref{fig:spectrum} are calculated with a drive strength that is often considered to be in the weak-drive regime. 
Nonetheless, it is perhaps more instructive to discuss in terms of an intermediate drive regime, $\Omd< \gcol$, prompted by the large losses that retain the system response in the weak-excitation regime.

\begin{figure}
    \centering
    \includegraphics[width=\linewidth]{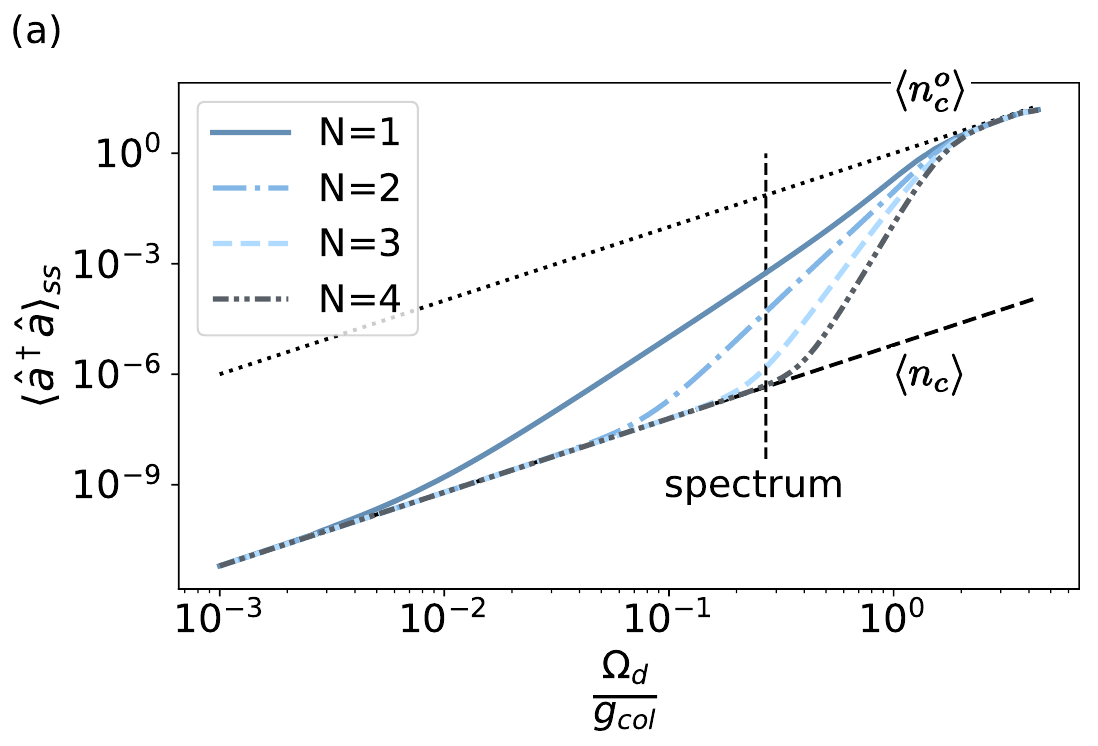}\\
    \includegraphics[width=\linewidth]{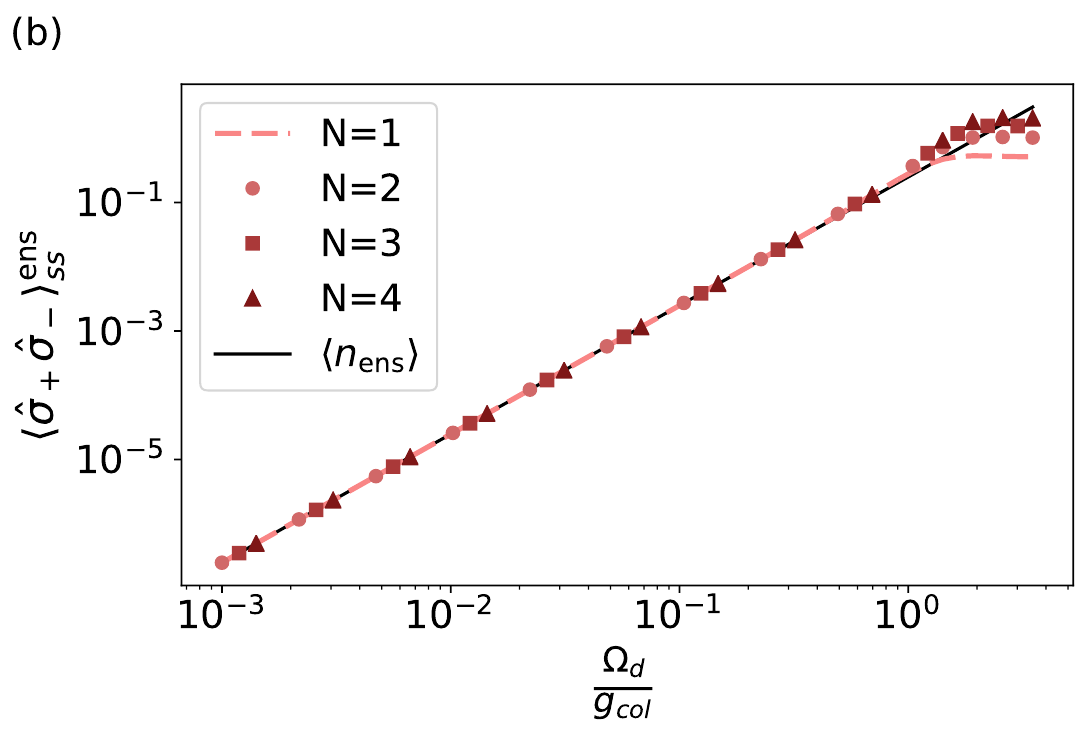}
    \caption{Log-log plots of the average steady-state populations as a function of the normalized drive strength $\Omega_d/\gcol$, calculated with the master equation for $N = 1-4$.
    (a) The cavity population shows an $N$-dependent transition through a nonlinear response regime for intermediate drive strengths between two linear asymptotes $\nccl$ (dashed black line) and $\ncbare$ (dotted black line). The dashed vertical line marks $\Om_d/g_{col}=0.25$, which was used for calculating the spectra in Fig.~\ref{fig:spectrum}. 
    (b) The total ensemble population for the same range of drive strengths as in panel (a). This population saturates at $N/2$ when the drive is strong.
    }
    \label{fig:ss(Om)}
\end{figure}

In Fig.~\ref{fig:ss(Om)}, we show the evolution of the steady-state cavity population $\langle \ad \a\rangle_{ss}$ and the steady-state total ensemble population $\langle \sp \sm \rangle^{\rm ens}_{ss}=\sum_1^N\langle\spi \smi\rangle_{ss}$ as a function of drive strengths from truly weak-drive conditions to strong drive ($\Omd>\gcol$). The populations are calculated for $N=1-4$ and are compared to the corresponding classical populations $\nccl$ and $\nenscl$ given by Eqs.~(\ref{eq:npl}) and (\ref{eq:nqe}). Intriguingly, as the drive strength increases, the results in Fig.~\ref{fig:ss(Om)}\figlink{(a)} show the cavity response entering into a nonlinear regime for each system size in turn. This effect is visible as an $N$-dependent break from the linear, classical response. Contrary to what naively could be expected in the weak-excitation regime, this implies that the few-level nature of a quantum-emitter ensemble plays an important role for the system dynamics, similar to the traditional saturation effect, already well before the system enters the strong drive regime. 

More specifically, \figpanel{fig:ss(Om)}{a} shows how the cavity response transition between two linear regimes at weak and strong driving, respectively. In the weak-drive regime, the cavity population follows the classical population $\nccl$ calculated with the CO model. Then, depending on the number of emitters $N$, a nonlinear regime is passed before the cavity again has a linear response described by an uncoupled driven damped harmonic oscillator with population $\ncbare=\Omd^2/\gamc^2$. 

Comparing the cavity population in \figpanel{fig:ss(Om)}{a} with the ensemble population in \figpanel{fig:ss(Om)}{b} shows that the nonlinear regime appears well before the emitter ensemble has saturated, i.e., when $\expval{\sp \sm}^{\rm ens}_{ss} \ll N/2$. It is not until strong drive conditions have been reached, $\Omd/\gcol > 1$, that the ensemble saturates and the cavity population approaches the response of an uncoupled driven harmonic oscillator.

A study of the slopes for the cavity population in the intermediate drive regime [\figpanel{fig:ss(Om)}{a}] shows the first evidence of the emitter-ensemble origin of the strongly nonlinear behavior. In linear response, the cavity population is expected to have a linear dependence on the driving intensity $I \propto \Omd^2$. This dependence is precisely what \figpanel{fig:ss(Om)}{a} shows for a sufficiently weak drive, where all systems have a slope of two in the log-log plot. If a multiphoton process of order $n$ is dominant, on the other hand, the cavity population would instead be proportional to the $n$th power of the driving intensity: $\expval{\ad \a}_{ss}\propto I^n \propto \Omd^{2n}$. Simple log-fits of the new slopes in \figpanel{fig:ss(Om)}{a} give the inclines $\sim 4,\, 6,\, 8$, and $10$ which would correspond to $2$-, $3$-, $4$-, and $5$-photon processes for the cases of $N=1,\,2,\,3$ and $4$ respectively. This indicates the dominance of $(N+1)$-photon processes facilitated by the ensemble of $N$ emitters.

\subsection{Multiphoton processes due to unconventional saturation effects}
\label{sec:EffectiveDrive}

To gain a better understanding of the observed dynamics, we study the system for weak external driving. By applying appropriate weak-drive approximations, the master equation in Eq.~(\ref{eq:ME}) simplifies to the same type of coupled equations of motion for the density-matrix elements~\cite{Carmichael_2008} as in the coupled-oscillator model described in \secref{sec:COmodel}. Therefore, more insight into the Tavis--Cummings dynamics in the weak-drive regime can be gained by observing the simple analytical solutions to the classical equations of motion presented in Eqs.~(\ref{eq:x0}) and (\ref{eq:xi}). 

This section includes two parts. In \secref{sec:destructive}, we show that the observed cancellation of cavity population in the weak-drive regime can be understood as a destructive interference between the external drive and the effective driving from the emitter ensemble due to the cavity-emitter coupling. Specifically, we formulate an effective drive on the cavity inspired by the classical equations of motion, which explain the cavity response in the weak-drive regime well.
In \secref{sec:breakdown}, we discuss how the breakdown of the destructive interference leads to the dominance of $(N+1)$-photon processes in the cavity response  at intermediate drive strengths. We also provide a simple phenomenological picture of the cavity response. In this picture, we neglect the ensemble and describe the observed $(N+1)$-photon process by multiphoton absorption events followed by cavity decay through single-photon processes.

\subsubsection{Destructive interference}
\label{sec:destructive}

In the coupled-oscillator model, an effective drive {\em on the cavity} can be defined by combining the external drive term with the coupling to the emitters, i.e. rearranging the terms in Eq.~(\ref{eq:x0}):
\begin{equation}
    \label{eq:Deff_cl_x}
    D_{\text{eff}}^{\text{cl}} = \Om_d \sqrt{2\hbar\omega_{0}}\cos(\omega_0 t) - \sum_i^N2g\omega_0 x_i.  
\end{equation}
The solution for $x_{i}$ is the real part of the ansatz $z_i=C_ie^{i\omega_d t}$ with the coefficient $C_i$ given by Eq.~(\ref{eq:Ci}). Inserting the solution for $x_i$ on resonance ($\omega_d = \omega_c = \omega_e \equiv \omega_0$) gives the effective drive 
\begin{equation}
    \label{eq:Deff_cl}
    D_{\text{eff}}^{\text{cl}}\Big|_{\text{res}} = \mleft(1-\frac{1}{1+\frac{\gamma_c \gamma_e}{4g^2_{col}}} \mright) \Om_d \sqrt{2\hbar\omega_{0}}\cos(\omega_0 t).
\end{equation}
Equation~(\ref{eq:Deff_cl}) shows that the external drive and the ensemble oscillator will interfere destructively. For $\gamma_c\gamma_e \ll  g^2_{col}$, the effective amplitude
\begin{equation}
    \label{eq:Om_eff}
    \Omega_{\text{eff}} = \mleft(1-\frac{1}{1+\frac{\gamma_c \gamma_e}{4g^2_{col}}} \mright) \Om_d \approx \frac{\gamma_c \gamma_e}{4g^2_{col}} \Om_d, 
\end{equation}
is much smaller than $\Omd$. This result explains the deep dip observed at resonance in the spectrum presented in \figpanel{fig:spectrum}{b}. Moreover, it elucidates the suppression of the coupled cavity population $\expval{n_c}$ compared to the uncoupled cavity population $\expval{n_c^o}$ shown in \figpanel{fig:ss(Om)}{a}.

A similar effective drive can be found in the Heisenberg picture for coupled quantum oscillators described by the position and momentum operators $\{\hat{x}_{c}, \hat{x}_{e,i}, \hat{p}_{c}, \hat{p}_{e,i}\}$. The effective quantum drive has the same form as \eqref{eq:Deff_cl_x}, but with the classical position variable $x_i$ replaced by the quantum operator $\hat{x}_{e,i}$. Since the eigenenergy spectrum of the emitter ensemble resembles a harmonic oscillator up to the same order of excitations as the number of emitters $N$, the validity of this coupled-oscillator picture is motivated. As such, the emitter ensemble behaves like a harmonic oscillator for weak excitation where higher-order terms are negligible. 
By employing this coupled-oscillator picture in the weak-excitation regime, a quantum analog to the classical effective drive described above can be formalized utilizing the properties of coherent states.

A coherently driven damped harmonic oscillator will also be in a coherent state. Hence, we can make a coherent-state approximation of the emitter ensemble to order $N$ in the weak excitation regime. In terms of Fock states, this coherent state can be written as
\begin{equation}
    \label{eq:alpha_ens}
    |\alphaens \rangle = e^{-\frac{|\alphaens|^2}{2}}\sum_{n=0}^N \frac{\alphaens^n}{\sqrt{n!}}|n\rangle. 
\end{equation}
The complex amplitude $\alphaens$ with $\abs{\alphaens}^2 \propto \expval{\sp \sm}^{\rm ens}_{ss}$ is defined by the emitter ensemble.

For this work, a rectangular `time-bin' temporal mode with duration $T$ is a sufficient description for the mode of the ensemble state in the weak-drive regime. This mode choice gives a simple expression for the complex amplitude:
\begin{equation}
	\alphaens = \Om_{\rm ens} T. 
\end{equation}
The time duration $T$ is a characteristic timescale set by the system. For the considered Tavis--Cummings system, the occupation of the ensemble is related to the cavity field through the collective coupling $\gcol$. Therefore, the natural choice of $T$ for the ensemble state is
\begin{equation}
    \label{eq:T}
	T = \frac{1}{\gcol}.
\end{equation}

To formulate a quantum analog to the classical effective drive, we have to relate the approximate coherent state for the ensemble to the classical external drive on the cavity. The relation can be found by considering an idealized laser for the external drive. The state of an idealized laser beam propagating through free space can be represented as a continuous-mode coherent state. 
 This continuous-mode coherent state can be partitioned into an infinite set of discrete-mode coherent states
\begin{equation}
    \label{eq:alpha_d}
    \ket{\alphad} = \exp^{-\frac{|\alphad|^2}{2}}\sum_{n=0}^\infty \frac{\alphad^n}{\sqrt{n!}}\ket{n}
\end{equation}
with amplitude $\alpha_d$.

The partitioning into discrete temporal modes can be performed with a large freedom of choice under the condition that the characteristic mode timescale $T_d$ and oscillator frequency $\omega_d$ obey  $\omega_d T_d \gg 1$. This freedom of mode choice is discussed in detail in \appref{app:Laser}, where we also give the partitioning into rectangular time bins as a specific example. Hence, in accordance with \appref{app:Laser}, we can also choose rectangular temporal modes for the external drive with a time duration $T_d$, which can be chosen arbitrarily as long as the condition $\omega_d T_d \gg 1$ is fulfilled. To compare the effects of the two sources of driving, the same choice of time duration must be made for the ensemble and the external drive. Thus, we take $T_d=T$, which gives the coherent state amplitude 
\begin{equation}
	\alphad = \Om_{d} T,
\end{equation}
with $T=1/\gcol$ as given by \eqref{eq:T} above.

Taken together with the coherent state approximation for the ensemble and the mode-matched partitioning of the laser beam, \eqref{eq:Deff_cl_x}
suggests that the cavity can be seen as driven by an effective coherent state in the linear regime. In terms of Fock states, this effective drive state can be written down as
\begin{equation}
    \label{eq:alpha_eff}
    \ket{\alphaeff} \approx \sum_{n=0}^N \frac{(\alphad-\alphaens)^n}{\sqrt{n!}} \ket{n}
\end{equation}
for small $\abs{\alphad}^2$ and $\abs{\alphaens}^2$.

For sufficiently weak drive, only lower-order Fock states ($n\leq N$) contribute notably to the scattering dynamics. In this regime, Eq.~(\ref{eq:Om_eff}) gives a classically derived analytical expression for the effective drive $\Omeff$, which agrees well with the numerical calculations using the master equation. By combining this picture of classical destructive interference with the idea of two mode-matched coherent states, it can be seen that Eq.~(\ref{eq:alpha_eff}) describes an effective, coherent drive on the cavity with amplitude
\begin{equation}
    \label{eq:alpha_eff_amp}
    \alphaeff = \Omeff T \equiv (\Omd -\Omens)T,
\end{equation}
where $T=1/\gcol$ is the characteristic timescale identified for the ensemble given in \eqref{eq:T}.
The validity of this choice of $T$ is confirmed by the good agreement between our analytical predictions using \eqref{eq:T} and the exact numerical calculations in the weak-drive regime presented in \figref{fig:ss(Om)}.


\subsubsection{Breakdown of destructive interference at intermediate drive strengths}
\label{sec:breakdown}

Under the weak drive conditions discussed so far, the classical and quantum models give the same result for the cavity field due to external driving. Nevertheless, there are distinct differences between the classical and quantum models that become clear in the effective-drive picture.
In the classical picture, the effective drive in \eqref{eq:Deff_cl} represents two harmonic fields acting on the cavity with opposite phase, which therefore cancels. Since both fields are harmonic, the relative amplitude $\Omeff$, written in \eqref{eq:Om_eff}, will not change when the drive strength increases.  Thus, the classical cavity population $\nccl$ maintains a linear dependence on the external drive in the strong-drive regime ($\Omega_d>\gcol)$. 

The emitter ensemble, on the other hand, only resembles a harmonic oscillator up to photon number $N$. This truncation of the harmonic spectrum is reflected in the effective coherent drive in  \eqref{eq:alpha_eff}, where the summation only goes to $N$. Hence, the destructive interference between the external drive and the ensemble breaks down at order $(N+1)$. However, the breakdown of the destructive interference is only visible when the ($N+1$)th state in the Fock-state expansion in \eqref{eq:alpha_d} has become significant. In \figpanel{fig:ss(Om)}{a}, the cavity response reveals this effect as $(N+1)$-photon processes for intermediate drive strengths. For even stronger drive, the emitter ensemble will saturate, and the uncoupled cavity response $\expval{n_c^o}$ will be approached.

The $(N+1)$-photon processes in the intermediate drive regime can be seen as a result of an \textit{unconventional saturation effect} where the emitter ensemble intermittently saturates on the cavity-emitter interaction timescale identified above.
This unconventional saturation is not visible in the spectrum since the emitter ensemble is still weakly populated and has \textit{not} saturated in the traditional sense, i.e., can not absorb more energy. Instead, the unconventional saturation effect can be described as the destructive interference between different excitation pathways which occurs on the characteristic timescale $T$ given in \eqref{eq:T}. 

\begin{figure*}
	    \includegraphics[width = 0.5\textwidth]{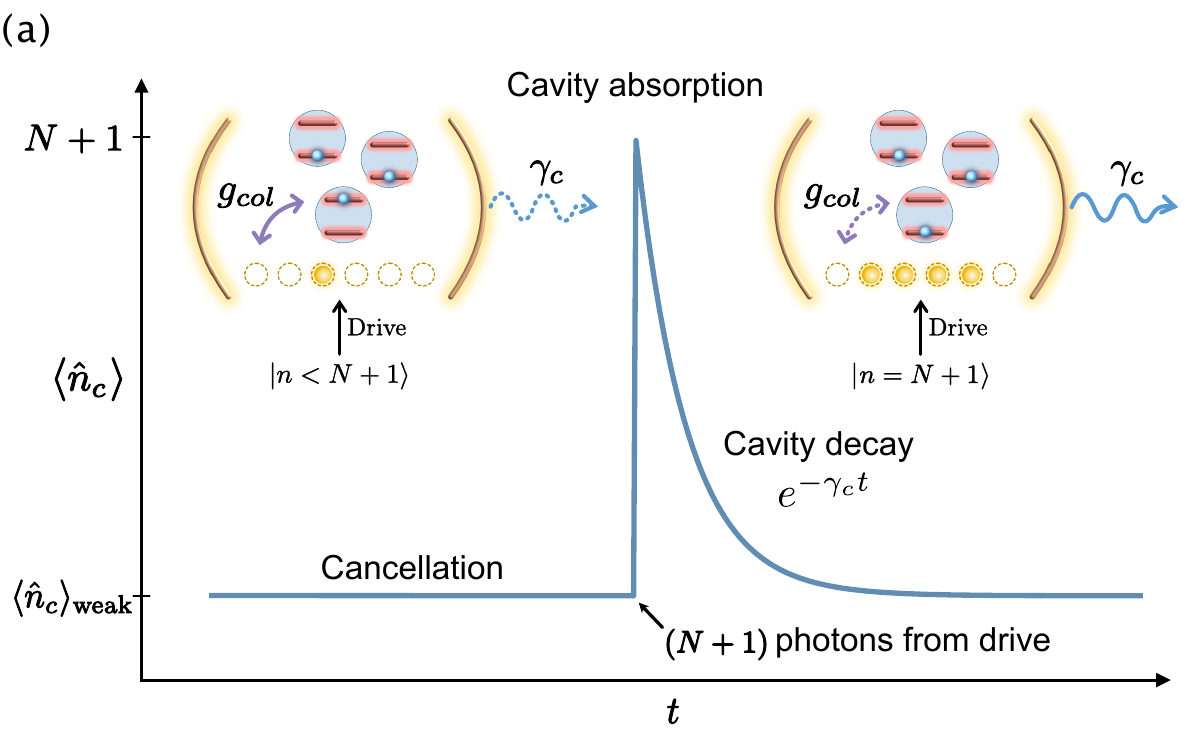}%
	    \includegraphics[width = 0.45\textwidth]{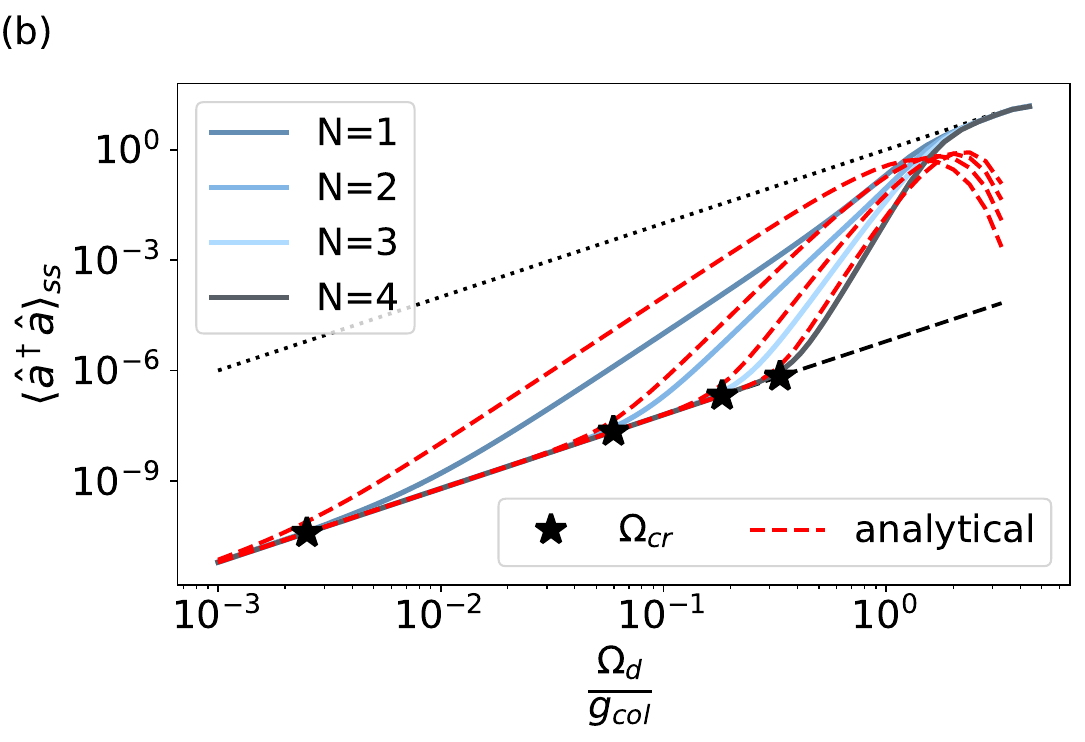}
    \caption{(a) The $(N+1)$-photon processes associated with the unconventional saturation effect can be seen as $(N+1)$-photon pulses driving the cavity when the destructive emitter-drive interference breaks down at order $(N+1)$. (b) The dynamics illustrated in panel (a) can be described with a phenomenological master equation which gives an analytical expression for the steady-state cavity population (red dashed curves). The analytical results are compared with the exact numerical calculations (solid blue curves). Black stars mark the critical drive for entering the unconventional saturation regime. }
    \label{fig:concept}
\end{figure*}

In a simplified picture, the unconventional saturation effect can be understood as a sequence of $(N+1)$-photon pulses driving the cavity due to the intermittent saturation of the destructive interference. In \figpanel{fig:concept}{a}, we provide a naive sketch illustrating one cycle of this $(N+1)$-photon process. The portrayed dynamics contain three distinct parts: (i) cancellation, (ii) $(N+1)-$photon absorption, and (iii) exponential decay. First, the cavity is mainly in the ground state with an average population $\expval{\hat{n}_c}_{\rm weak}$ due to the emitter-drive interference, which cancels the cavity population. However, when $(N+1)$ photons arrive from the drive, the destructive interference breaks down. On the timescale of the cavity-emitter interaction, the emitters intermittently saturate, which leads to direct absorption of the $(N+1)$-photon state in the cavity. Following the absorption event is exponential decay, where the $(N+1)$-photons leak out of the cavity photon-by-photon.

The dynamics illustrated in \figpanel{fig:concept}{a} can be modeled with a simple phenomenological master equation for the probabilities $P_n(t)$ of occupying the $n$th Fock state in the cavity.
Despite its simplicity, considering only cavity processes, this master equation qualitatively captures the unconventional saturation effect for drive strengths approaching the collective coupling $ \gcol$. The details of this approach are shown in \appref{app:multiphoton_pulses}. Here we present the main results.

Most importantly, the phenomenological master equation presented in \appref{app:multiphoton_pulses} provides analytical solutions to the time-dependent probabilities $P_n(t)$. These solutions allow us to make an analytical prediction of the steady-state cavity population by calculating the time-averaged contribution from having a stream of $(N+1)$-photon pulses driving the cavity due to the unconventional saturation effect. The steady-state cavity population in this naive picture can be found as
\begin{equation}
	\label{eq:nss_analytical}
	\expval{\hat{n}_c}_{ss} = \ncweak+ \frac{(N+1)P_{N+1}^T}{T \gamma_c},
\end{equation}
where $\ncweak$ is the suppressed cavity population due to the coupling to the ensemble and $P_{N+1}^T$ is the probability of having $(N+1)$ photons in the external drive during the time $T$ given by \eqref{eq:T}. Here we have also introduced the notation $\expval{\hat{n}_c}_{ss}=\expval{\ad \a}_{ss}$ for the steady-state cavity population.

An expression for $\ncweak$ can be found by employing the coupled oscillator model explained above.
In \figpanel{fig:ss(Om)}{a}, we have already seen that the steady-state cavity population is well described by the classical result $\expval{n_c}$ for a sufficiently weak drive. Therefore, Eqs.~(\ref{eq:C0}) and (\ref{eq:npl}) give an expression for $\ncweak$ which accurately predicts the cavity population in the weak-drive regime. On resonance, this expression is
\begin{equation}
    \label{eq:nc}
    \ncweak = \frac{\Omd^2\gamma_e^2}{16\gcol^4}\frac{1}{\mleft(1+\frac{\gamma_c\gamma_e}{4\gcol^2}\mright)^2}.
\end{equation}
The probability $P_{N+1}^T$ is given by the Poisson distribution for the external drive with discrete-mode amplitude $\alpha_d  = \Omd T$,
\begin{equation}
	\label{eq:poisson_drive}
	P_{N+1}^T = \exp^{-\abs{\Omd T}^2}\frac{\abs{\Omd T}^{2(N+1)}}{(N+1)!}.
\end{equation}

Figure~\ref{fig:concept}\figlink{(b)} shows the steady-state cavity population for $N=1-4$ emitters obtained with \eqref{eq:nss_analytical} (red dashed curves). The analytical results are compared with the exact numerical calculations using the master equation in \eqref{eq:ME} (blue solid curves).
The comparison shows that the simplified dynamics presented in \figpanel{fig:concept}{a}, leading to the analytical prediction for $\expval{\hat{n}_c}_{ss}$ given in \eqref{eq:nss_analytical}, capture the unconventional saturation effect qualitatively.  It can be seen that \eqref{eq:nss_analytical} accurately captures both the weak-drive behavior and the dominance of $(N+1)$-photon processes in the intermediate drive regime. The simple picture of $(N+1)$-photon pulses arising from the breakdown of destructive interference at order $(N+1)$ can thus give a qualitative intuition for the unconventional saturation effect.

The analytical prediction overestimates the photon number. However, this is not surprising since the phenomenological model employed to derive \eqref{eq:nss_analytical} entirely neglects all effects from the coupling to the ensemble beyond the cancellation effect. For example, the possibility of excitation transfer to the ensemble during the decay process is completely overlooked. Nevertheless, the simplified model described in this section is a good tool that can be used to gain an intuition about the $(N+1)$-photon processes associated with the unconventional saturation effect.

Equation (\ref{eq:nss_analytical}) can also be extended to account for higher-order photon pulses ($n>N+1$). In that case, the last term becomes a sum of the contributions. See \appref{app:multiphoton_pulses} for details. Since the external drive is coherent, the probabilities $P_n^T$ for the higher-order photon states ($n\geq N+1$) follow a Poisson distribution. Hence, the $(N+1)$-photon pulses contribute the most to the cavity response at intermediate drive strengths $(\Omd < \gcol)$. Including higher-order photon pulses in this simplified picture will, therefore, not qualitatively change the cavity response in this regime.  In \appref{app:multiphoton_pulses}, we show a calculation including the contribution of photon pulses up to order $N+5$, which confirms our argument above.

\subsection{Critical drive strength}
\label{sec:CriticalDrive}
The effective-drive picture, and the breakdown of the destructive interference discussed above, can also be used to write down a condition for entering into the nonlinear regime. 
Under the assumption that the emitter ensemble behaves as a driven harmonic oscillator up to order $N$, we would expect the cavity response to enter the nonlinear regime when the missing $(N+1)$th term in the coherent-state approximation $\ket{\alphaens}$ becomes comparable to the cavity population. From the coupled-oscillator perspective, this condition is easy to understand. That is, up to $N$ excitations, the system behaves classically, and the ensemble can interfere destructively to cancel out excitation of the cavity.
On the other hand, when the ensemble fails to interfere destructively due to its few-level spectrum, the cavity population becomes comparable to the missing $(N+1)$th term in $\ket{\alphaens}$. Formally, this condition for the critical drive can be written down as
\begin{equation}
    \label{eq:conditon}
    \expval{ \hat{n}_c}_{ss} = (N+1) \text{P}_{\alphaens}(N+1).
\end{equation}
Here $\text{P}_{\alpha}(n)\approx \frac{|\alpha|^{2n}}{n!}$ is the Poisson probability distribution for finding $n$ excitations in the coherent state when $|\alpha|^2\ll1$ and the factor $(N+1)$ comes from having $(N+1)$ excitations with probability $\text{P}_{\alpha}(n)$.

Since we are approaching the nonlinear regime from the weak-drive regime, we can take $\expval{\hat{n}_c}_{ss} = \ncweak$ and use the expression for $\ncweak$ given in \eqref{eq:nc}.
%
%
We can also find an expression for the effective ensemble drive amplitude $\alphaens = \Omens T$ using $\Omeff$ in \eqref{eq:Om_eff}:
\begin{equation}
    \label{eq:alpha_ens_amp}
    \alphaens = \Omens T = \frac{\Omd}{1+\frac{\gamma_c \gamma_e}{4\gcol^2}}T.
\end{equation}
The condition in Eq.~(\ref{eq:conditon}) thus becomes
\begin{equation}
    \ncweak = (N+1)\frac{\abs{\alphaens}^{2(N+1)}}{(N+1)!},
\end{equation}
which gives the expression
\begin{equation}
	\Omega_{\rm cr}(N) =  \left( \frac{N! \gamma_e^2 g_{\rm col}^{2(N-1)}}{16} \right)^{\frac{1}{2N}} \left( 1+\frac{1}{C}\right)
	\label{eq:OmB}
\end{equation}
for the critical drive strength $\OmB$ that indicates the onset of the unconventional saturation regime.
\begin{figure*}
    \includegraphics[width=\linewidth]{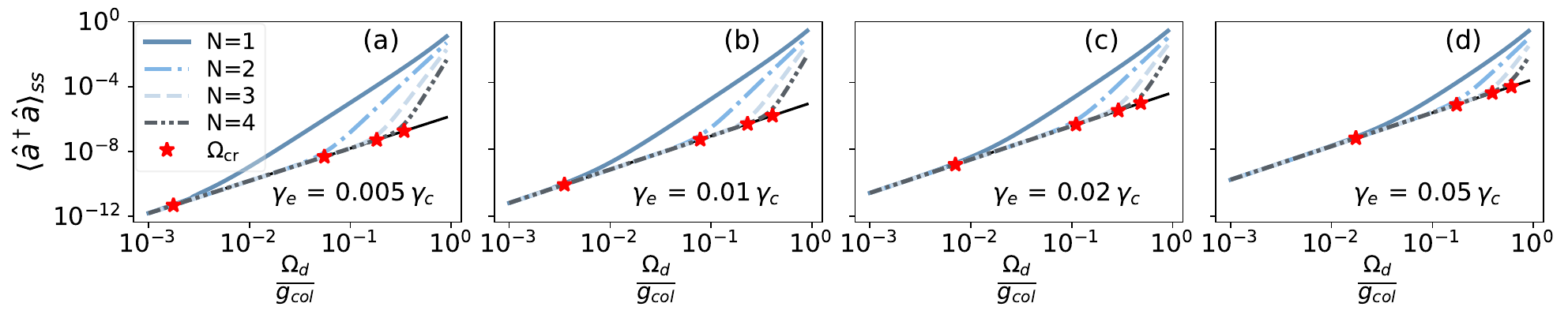} \\
    \includegraphics[width=\linewidth]{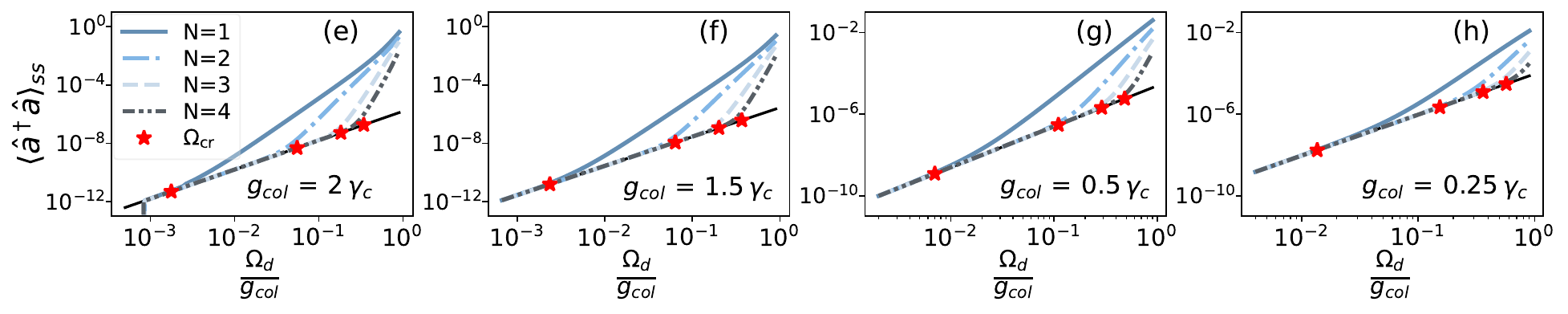}
    \caption{Calculations of the steady-state cavity population for different [(a),(b),(c),(d)] emitter decay rates $\game$ and [(e),(f),(g),(h)] collective interaction strengths $\gcol$. The remaining parameters were held fixed, using the same values as in Fig.~\ref{fig:ss(Om)}. The red stars mark the derived analytical expression for the critical drive $\OmB$. As can be seen, the analytical results predict exceedingly well the onset of the nonlinear regime.}
    \label{fig:OmB}
\end{figure*}
 
As can be seen in Fig.~\ref{fig:OmB}, the calculation of the critical drive with Eq.~(\ref{eq:OmB}) predicts very well the onset of the nonlinear regime. Figures~\ref{fig:OmB}\figlink{(a)-(d)} show the steady-state cavity response and the calculated $\OmB$ for a wide range of emitter decay rates $\game$, whereas Figs.~\ref{fig:OmB}\figlink{(e)-(h)} show the same for several different coupling strengths $\gcol$. In all panels, the analytically calculated $\OmB(N)$, marked with red stars, lie very close to the beginning of the nonlinear regime. Thus, we have not only found a new intriguing regime for performing quantum nonlinearity measurements, but we can also, with high accuracy, predict its onset for a wide range of parameters.

\subsection{Figure of merit}
\label{sec:FOM}
So far, we have discussed unconventional saturation as an effect of the destructive interference (or the competition of) two distinct excitation pathways: cavity-drive and cavity-emitter. That interesting quantum effects can arise in dissipative Tavis--Cummings type systems due to quantum interference has already been shown with the so-called unconventional photon blockade effect~\cite{Radulaski_2017,Bamba_2011,You_2020,Snijders_2018,Flayac_2017,Saez-Blazquez_2017,Saez-Blazquez_2018}. The unconventional photon blockade, however, is observed for weak excitation and strong-coupling conditions. The unconventional saturation effect observed here, on the other hand, is instead present for intermediate drive strengths and appears for resonant driving in a parameter regime where unconventional photon blockade is absent.

It turns out that a good figure of merit for unconventional saturation is the cooperativity
\begin{equation}
    \label{eq:Coop}
    C \equiv \frac{4\gcol^2}{\gamc\game}.
\end{equation}
Why \eqref{eq:Coop} is a good figure of merit can be seen by studying the induced transparency (reduced cavity population) on resonance due to the effective drive in the classical case. By taking the ratio between the classically derived coupled- and uncoupled-cavity populations, it can be seen that the cavity response will be suppressed with a factor depending on $C$:
\begin{equation}
    \label{eq:FOM}
    \frac{\expval{n_c}}{\expval{n_c^o}} = \frac{1}{\mleft(\coop+1\mright)^2}.
\end{equation}
To arrive at this relation, we have used the expression in \eqref{eq:nc} for $\expval{n_c}$, found by solving the classical coupled-oscillator equations of motion. A similar calculation for an uncoupled cavity, driven by the same external drive on resonance gives 
\begin{equation}
    \label{eq:nc_o}
    \expval{n_c^o} =  \frac{\Omd^2}{\gamma_c^2},
\end{equation}
as already mentioned above.

In \figpanel{fig:ss(Om)}{a}, it can be seen that \eqref{eq:FOM} governs the region in which the unconventional saturation effect can be observed. For small cooperativities, i.e., $C\sim 1$ or smaller, the suppression of the cavity response due to the interaction with the emitter ensemble is too small for observing unconventional saturation. However, for $C \gtrsim 10$ the unconventional saturation effect starts to become clearly visible, and (as would be expected) it grows more distinct for increasing $C$. In \appref{app:coop}, additional simulations that show how the cavity response changes with the cooperativity can be found.

\subsection{Suppression of the cavity response}
\label{sec:suppression}
Equation~(\ref{eq:FOM}) in \secref{sec:FOM} shows a classically derived expression for the suppression of the coupled-cavity response in the linear weak-drive regime. This result, together with the effective drive $\ket{\alphaeff}$ found in Eq.~(\ref{eq:alpha_eff}), underlines the expectation of a transition in the cavity response from a coupled coherent state to an uncoupled coherent state as the drive is increased. And indeed, this is what we see in \figpanel{fig:ss(Om)}{a}. In \secref{sec:EffectiveDrive}, we also identified the timescale $T=1/g_{\rm col}$, which defines the characteristic time for unconventional saturation. This timescale provides a simple relationship between the cavity decay time $T_c=1/\gamma_c$ and the effective (suppressed) drive $\alpha_{\rm eff}$, which is easily found by rewriting the expression for $\expval{n_c}$ in terms of the effective drive amplitude $\Omeff$ in \eqref{eq:Om_eff} and employing the relations for $\alphaens$ in \eqref{eq:alpha_eff_amp}:
\begin{align}
    \label{eq:nc_timescales}
    \expval{n_c} & = \frac{\Omeff^2}{\gamma_c^2} = \left(\frac{T_c}{T}\right)^2\abs{\alphaeff}^2.
\end{align}

The arguments above explain the two asymptotical behaviours observed and demonstrate the competition of timescales causing the unconventional saturation effect. In the following section, 
we show that the populations of the reduced density matrix for the cavity lie between the Poissonnian distributions for the two asymptotical coherent states. Moreover, we discuss how the destructive interference with the $N$-emitter ensemble appears in the populations and how they evolve as the drive increases.

\subsubsection{Suppression of populations}
As already noted, the nonlinear cavity response will appear in the transition between the response of a coupled cavity and that of an uncoupled cavity. In the two linear regimes, at weak and strong drive, respectively, the cavity will be described by the two coherent states with amplitudes $\alpha_c$ and $\alpha_c^o$. To find the amplitudes, we can use the well-known property $\abs{\alpha}^2=\expval{\hat{n}}$ for coherent states and use the classical results derived above. Thus, we find
\begin{equation}
    \abs{\alpha_c}^2 =\frac{\Omeff^2}{\gamc^2}
\end{equation}
for the coupled coherent state and
\begin{equation}
    \abs{\alpha^o_c}^2 = \frac{\Omd^2}{\gamc^2}
\end{equation}
for the uncoupled coherent state.

\begin{figure*}
	    \includegraphics[width = 0.5\textwidth]{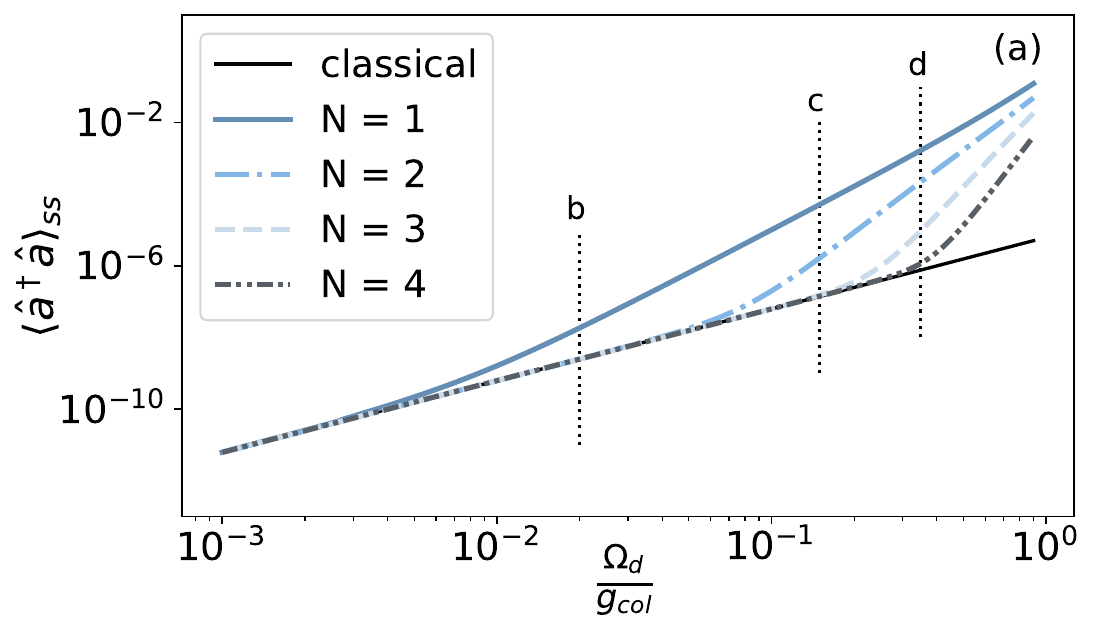}%
	    \includegraphics[width = 0.5\textwidth]{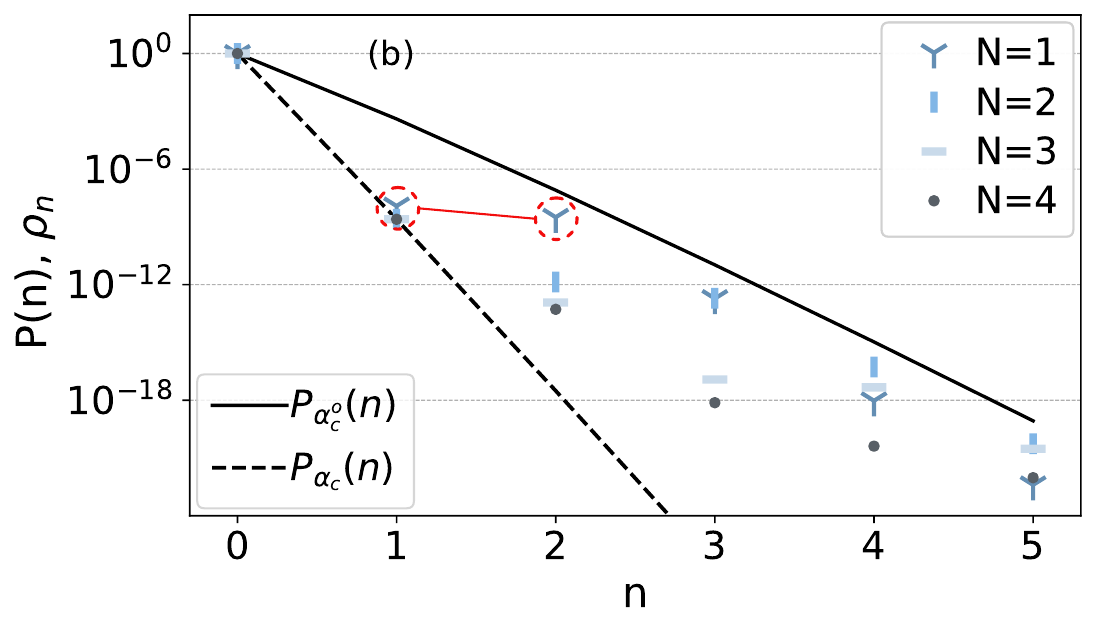}\\
	    \includegraphics[width = 0.5\textwidth]{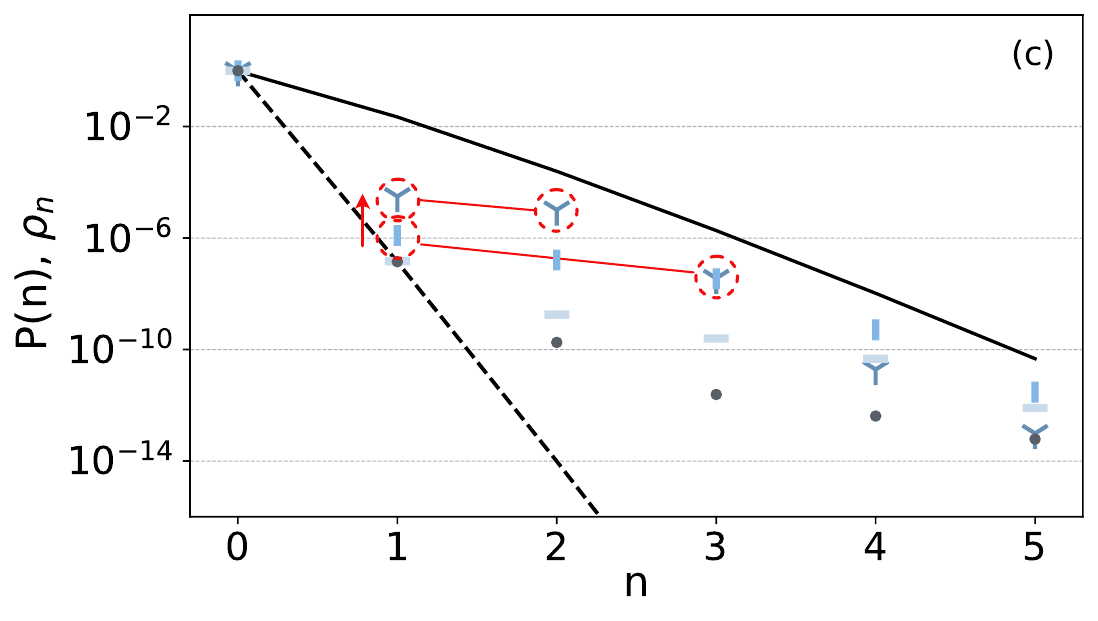}%
	    \includegraphics[width = 0.5\textwidth]{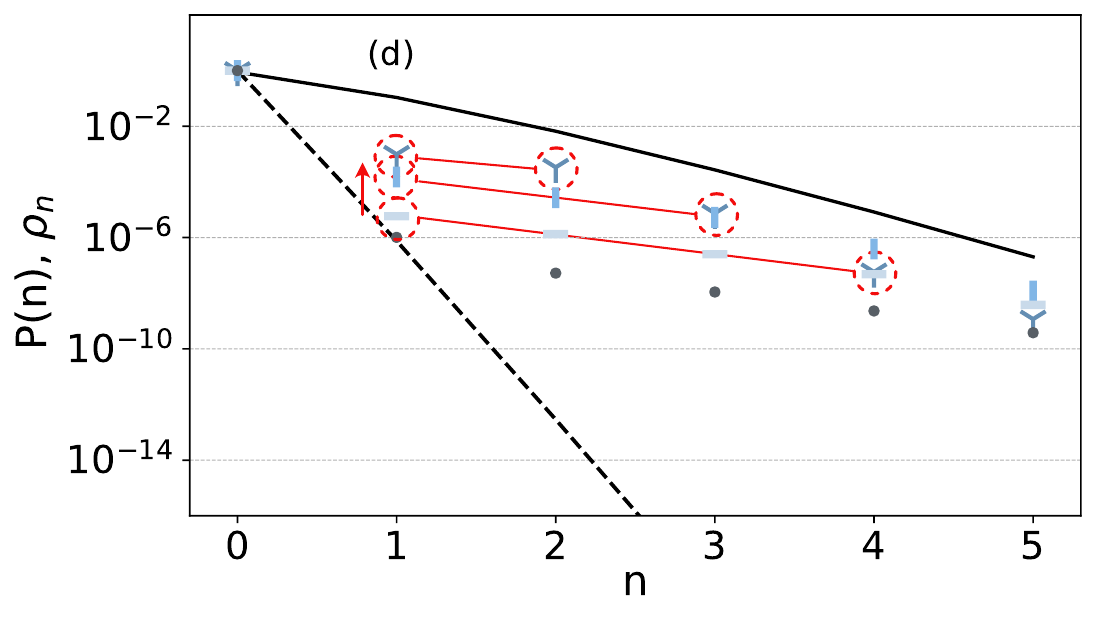}%
    \caption{Visualization of the evolution of [(b),(c),(d)] the multiphoton populations $\rho_{n}$ as the drive strength $\Om_d$ is increased; the total population is shown in panel (a). The dashed and solid lines mark the Poisson distributions $P(n)$ calculated with the coherent-state amplitudes corresponding to a coupled, $|\alpha_c|^2=\Omeff^2/\gamc^2$, and uncoupled, $|\alpha_c^o|^2=\Om_d^2/\gamma_c^2$, cavity, respectively. As can be seen for all system sizes $N = 1$--$4$, the emitter ensemble behaves as a true coherent state in the coupled system only up to first order. Then, for increasing drive strength, each system \textit{saturates} in turn, visible as the higher-order photon states approach the uncoupled-cavity distribution. The red circles illustrate when $\rho_{N+1}$ have become comparable to $\rho_{1}$, i.e., $\rho_1 \sim (N+1)\rho_{N+1}$}, which is a signature of the unconventional saturation regime.
    \label{fig:dm_elements}
\end{figure*}

To gain information on the state inside the cavity, we examine the populations $\rho_n \equiv \expval{\rhoh_c}{n}$ of the reduced density matrix $\rhoh_c \equiv \Tr_{\rm ens}\{\rhoh_{ss} \}$. In \figref{fig:dm_elements}, the cavity populations for $N = 1,\, 2,\, 3$, and $4$ emitters (symbols) are plotted against the Poisson distributions for the coherent states with amplitudes $\alpha_c$ (dashed line) and $\alpha_c^o$ (solid line) for three different drive strengths as indicated (dashed vertical lines) in \figpanel{fig:dm_elements}{b}.

Because of the destructive interference, we expect the populations to approach the coupled Poisson distribution, in the weak-drive regime. However, due to the truncation of the ensemble spectra at $N$ excitations, the higher-order photon-state `tail' will be pulled towards the uncoupled Poisson distribution. This effect can be viewed as a sequential flow of excitation from the $\rho_{n\geq N+1}$ populations, that are not affected by the cancellation, to the $\rho_{n<N+1}$ populations via decay processes.

When the system is \textit{not} in the unconventional saturation regime, we expect the distribution of the populations to be `Poisson-like' ($\rho_0\gg \rho_1 \gg \rho_2 ...$) except for the truncation-induced cross-over explained above.
On the other hand, when the system \textit{is} in the unconventional saturation regime, the breakdown of the destructive interference at order $N+1$ will facilitate direct $(N+1)$-photon excitation of the cavity; see \figpanel{fig:concept}{a}. Following the absorption is cavity decay, which is again a single-photon process.  Thus, in the unconventional saturation regime, we expect the $\rho_{N+1}$ population to be comparable to the lower-order populations $\rho_{n<N+1}$, that is, $(N+1)\rho_{N+1} \approx N\rho_{N} \approx ... \approx \rho_1$. This picture agrees well with the breakdown of destructive interference discussed in \secref{sec:breakdown}. In \appref{app:multiphoton_pulses}, we additionally show that the simple phenomenological master equation discussed in \secref{sec:breakdown} also predicts the populations qualitatively.

\subsection{Relation to exciton-induced transparency}
\label{sec:QuantumInterference}

\begin{figure*}
    \centering
    \includegraphics[width=\textwidth]{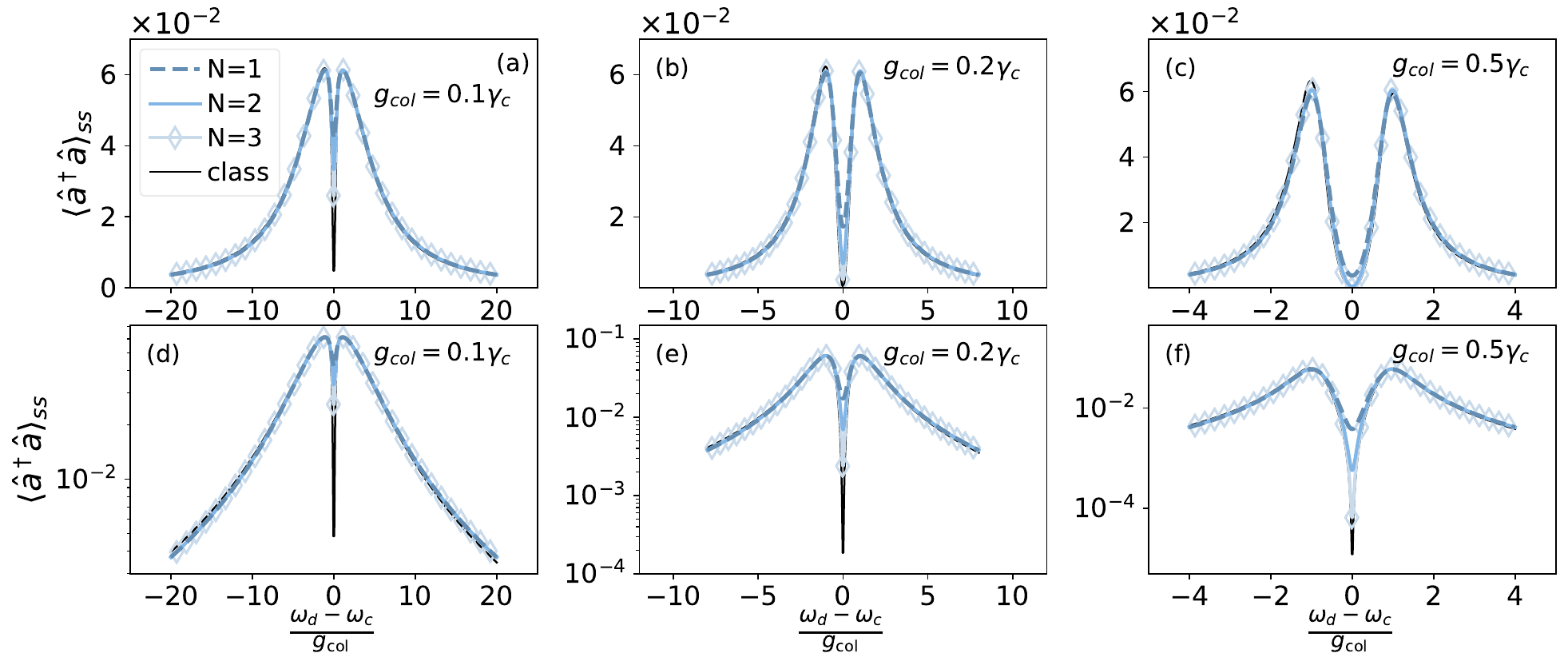}\\
    \caption{Spectra of the average cavity population $\langle \ad \a \rangle$ for $N= 1, \, 2$, and 3 emitters plotted against the classical analogue $\langle n_c\rangle$ on a [(a),(b),(c)] linear and [(d),(e),(f)] logarithmic scale. As is clearly seen throughout the log-log plots, the quantum interference effect is present for a wide range of interaction strengths below and at the strong-coupling limit $\gcol \geq \gamma_c/2$.}
    \label{fig:EIT}
\end{figure*}

The underlying destructive interference effect, giving rise to the observed suppression of the cavity population on resonance, can in the weak drive regime be understood as a classical analogue to electromagnetically induce transparency (EIT)~\cite{Alzar_1995}. Indeed, the coupled set of equations presented in Eqs.~(\ref{eq:x0}) and (\ref{eq:xi}) is the same as those used for modeling classical EIT. EIT is more commonly discussed in systems with more moderate light-matter coupling than illustrated in Fig.~\ref{fig:spectrum}. In Fig.~\ref{fig:EIT}, the cavity population spectrum is plotted for $N= 1, \,2$, and 3 emitters together with the classical analogue for increasing interaction strength $\gcol$. At the top, in \figpanel{fig:EIT}{a} and \figlink{(b)}, the typical EIT regime with a rather moderate interaction strength $\gcol< \gamc/2$ is shown. As is clearly visible throughout the log-log plots in the right panel, the $N$-dependent quantum interference effect at resonance persists for a wide range of interaction strengths. 

For a system with one emitter, a semiclassical analog to classical EIT, treating the emitter quantum mechanically and the cavity classically, has already been referred to as \textit{'exciton-induced transparency'} (ExIT). ExIT has been discussed within the context of plasmon-exciton coupling~\cite{Artuso_2010,Ridolfo_2010,Pelton_2019} and demonstrated with a plasmonic nano-cavity and quantum-dot system~\cite{Wu_2010}. 
The saturation of the ExIT effect has been discussed in Ref.~\cite{Shah_2013}, although the exact power dependence was not analyzed. The regime with dominant two-photon processes for this single emitter system was thus not identified. However, the systems exhibiting ExIT should indeed be suitable for demonstrating unconventional saturation.

\subsection{Relation to dressed-state polarisation}
\label{sec:dressed-state-polarisation}

In \figpanel{fig:ss(Om)}{a}\, two critical points can be identified. The first point, at weak drive, marks the onset of the unconventional saturation effect, and it depends strongly on the emitter number, as discussed in \secref{sec:CriticalDrive}. The second point, at strong drive, marks the conventional saturation of the emitter ensemble, as can be evidenced by also studying the steady-state ensemble population in \figpanel{fig:ss(Om)}{b}. This point is related to the collective coupling strength. Thus, it is primarily the same for all system sizes with the same collective coupling. 

In a resonantly driven Jaynes--Cummings model ($N=1$), the critical point at strong drive has additionally been identified as a first-order phase transition for a strongly coupled system~\cite{Alsing_1991,Alsing_1992,Carmichael_2015}. Due to the strong cavity-emitter coupling, the Jaynes--Cummings system can be seen as an atom-cavity \textit{`molecule'} that becomes dressed by the resonant external drive. Beyond the critical drive strength, which in our parameters translates to $\Omega_d/g_{\rm col}=1$, the system exhibits phase bimodality in the steady state. This effect is known as \textit{dressed-state polarisation} and has been analysed in detail in Refs.~\cite{Alsing_1991, Alsing_1992}. The dressed-state polarisation effect has also been connected to the breakdown of the photon-blockade effect~\cite{Carmichael_2015}, and it has been studied with mean-field theory in the limit $N\to \infty$ \cite{gutierrez-jauregui_dissipative_2018}. 

Unlike the unconventional saturation effect, the dressed-state polarisation effect is a strong-coupling effect in the strong drive regime, that relies on climbing the dressed Jaynes--Cummings ladder. 
The unconventional saturation effect, on the other hand, is present in, but not limited to, the strong-coupling regime. Instead, as discussed in \secref{sec:FOM}, the relevant figure of merit for unconventional saturation is large cooperativity. By virtue of the single-emitter decay rate in the denominator in \eqref{eq:Coop}, a large cooperativity can be obtained even though the cavity loss is large (i.e., weak coupling) if the emitter decay is small. Thus, unconventional saturation can be present in systems which are prohibited from displaying dressed-state polarisation in the strong drive regime due to the large cavity dissipation. 
Moreover, in contrast to dressed-state polarisation, the unconventional saturation effect appear at intermediate drive strengths, well before the critical point, $\Omega_d/g_{\rm col}=1$, marking the phase transition to the dressed-state polarisation phase in the resonantly driven Janyes--Cummings model.

\subsection{Limitations of the model}
\label{sec:limitations}
With the aim of formulating the simplest model that predicts the main physical phenomena arising with small emitter ensembles coupled to a cavity in an environment, the
master equation employed in this work is derived under two important conditions: (i) the cavity and emitters interact with the environment independently of the cavity-emitter coupling, and (ii) each emitter interacts with local environments. This approach follows a criterion of simplicity and is appealing since it keeps the discussion of the physics simple. 
 Below, we discuss the limitations of the conditions (i) and (ii). Moreover, we also discuss how small changes to these conditions do not change the effect presented in this work. 
 
In the following,  we refer to the eigenstates of the coupled cavity-emitter system as dressed states, and we refer to the eigenstates of the cavity and emitter subsystems as bare states.

\subsubsection{The rotating-wave approximation}

Condition (i) states that the dissipative part of the master equation is derived without taking into account the cavity-emitter interaction. 
The dissipators [$\mathcal{D}_{\a}$ and $\mathcal{D}_{\sm}$ in \eqref{eq:ME}] obtained through this method will induce transitions between the bare cavity and emitter states. The master equation obtained in this fashion can accurately describe many cavity-QED experiments in regimes where the RWA can safely be applied.

Generally, the RWA is justified for systems with  $g_{col}/\omega_c <0.1$, which is the conventional limit for the ultra-strong coupling (USC) regime~\cite{Kockum2019}.  In this article, we demonstrate an unconventional saturation effect with a set of parameters giving $g_{col}/\omega_c=0.03$, which is clearly below the limit of USC. Moreover, as discussed in \secref{sec:FOM}, the effect is also present for lower $g_{col}$ provided that the cooperativity $C$ is large enough. Thus, the RWA should, in general, be justified in the regimes where the unconventional saturation effect typically appears. 

However, the limit $g_{col}/\omega_c <0.1$ is merely a historical convention, not a strict boundary for when the RWA can safely be used. Therefore, when a high degree of accuracy is demanded, e.g., in the case of low photon numbers, the validity of the RWA may have to be reevaluated. When the RWA is not applied, the system Hamiltonian is no longer excitation-number conserving. The mixing of the bare ground state with higher-order states leads to virtual excitations in the dressed ground state (see, e.g., Ref.~\cite{Kockum2019} for a review on USC discussing virtual excitations in the ground state). For cavity-emitter systems that are not ultra-strongly coupled, the effects from the mixing of states containing different numbers of excitations are typically negligible, but the effect demonstrated in this manuscript appears at very low excitation numbers. The change of the ground state when the RWA is not applied could therefore be important.

The retraction of the RWA requires proper adjustments to the master equation in \eqref{eq:ME}, which would otherwise give unphysical excitation out of the coupled ground state even without driving at zero temperature. In \appref{app:RWA}, 
we provide a master equation that induces transitions between the dressed states and accurately brings the system to the dressed ground state, including the counter-rotating terms in the Hamiltonian. Employing this master equation, we find that the counter-rotating terms do not qualitatively affect the unconventional saturation effect. For a comparison of the results obtained with the different master equations, see \figref{fig:ME_comparison} in \appref{app:RWA}.

\subsubsection{Local emitter environments}

The second condition (ii) neglects any effects of collective interaction between the emitter ensemble and the environment. For example, in optical bistability, which also appears within the drive-dissipative Tavis--Cummings model, the individual dissipation approach is the textbook procedure~\cite{Carmichael_2008_ch15,gardiner_quantum_2004_ch9}.  Optical bistability was also recently studied with the individual-emitter-decay master equation in a low-photon-density regime~\cite{shirai_optical_2018} similar to ours. Thus, individual emitter decay seems to be an adequate method to model the central physical phenomena in the driven-dissipative Tavis--Cummings model. The connections between optical bistability and unconventional saturation in this regime also remain an interesting question for future study.

Nevertheless, the work of Dicke~\cite{Dicke_1954} highlighted the importance of collective dissipation for atoms contained in a small volume compared to the wavelength, which interacts with a single mode in free space. In that setting, the collectiveness of the interaction gives rise to the well-known superradiance effect~\cite{gross_superradiance_1982}. Since then, the concept of collective decay has been transferred to cavity-QED systems, where quantum emitters are confined to a small volume inside the cavity.

In this work, all emitters interact uniformly with the cavity mode. This approximation is strictly only valid for emitter ensembles that are localized in a small volume compared to the mode wavelength, but serves as a good first approximation for many other experimental configurations. Because of small emitter confinement, the collectiveness of the interaction with the environment could be non-negligible for accurately describing the system dynamics.
However, with the current set of parameters that show the unconventional saturation effect, the interaction between the emitters and the environment is weak and much smaller than the cavity dissipation ($\gamma_e \ll \gamma_c$). 
Therefore, the main decay channel for the emitters is through the cavity.  Under these circumstances, where emitter emission is a rare event mainly shielded by the cavity, it is not unreasonable to assume that the remaining dissipation from the emitters will appear through local decay channels. Another situation where individual emitter dissipation would be a good approximation is for localized emitters, e.g., on a substrate.

Another important aspect is the treatment of only a few emitters. 
In the few-emitter regime, the difference between collective and individual dissipation for the emitters is minimal. So, a collective-dissipation picture does not qualitatively change the unconventional saturation effect; see  \figref{fig:ME_comparison} in \appref{app:RWA}. 
The exact limits to the validity of individual versus collective dissipation is an interesting question that would require further theoretical work.

%
%

\section{Conclusions}
\label{sec:Conclusions}
In this work, the stationary response from a coherently driven cavity coupled to an ensemble of $N$ quantum emitters, described by the Tavis--Cummings model, has been studied. The steady-state density matrix was calculated numerically using a master-equation approach without making the frequently applied weak-drive approximation. Additionally, a classical coupled-oscillator model was applied to give analytical insight into the dynamics in the linear regimes. 

For resonant drive frequency and intermediate drive strength, our results show strongly $N$-dependent nonlinear scattering. Specifically, we see the dominance of $(N+1)$-photon processes in the nonlinear regime of the cavity response when it couples to an ensemble of size $N$. In contrast to observing Rabi splitting in the spectrum, this effect clearly differentiates between different ensemble sizes $N$ with the same collective interaction strength $\gcol$. 

Exploiting analytical results from a classical coupled-oscillator model, and properties of coherent states, we found that the origin of this effect could be explained by the destructive interference between the ensemble and the coherent drive up to the order $N$. Thus, the ensemble behaves as a saturable mirror that can only reflect photon states up to order $N$. This unconventional saturation effect occurs due to a competition of interaction rates and arises for weak ensemble population, well before traditional saturation. We also derived an analytical expression for the critical drive $\OmB$ that to good accuracy predicts the onset of the nonlinear regime. Moreover, we find that a basic condition for the observed unconventional saturation effect is large cooperativity $C$. This condition can be met without the requirement of strong coupling, if the decay rates of the emitters are not too large.

The observed effect implies a simple continuous-wave method that could characterize dissipative cavity-emitter systems where the number of quantum emitters is unknown. The $N$-dependent interference effect and the resulting $(N+1)$-photon processes in the cavity response could also be exploited for photon filtering. Thus, our results show great promise for the use of dissipative cavity-few-emitter systems for quantum state engineering. For this, further theoretical work that investigates the specific output state for different input states would be of interest.  

Lastly, we note that a first-order expansion of the Holstein-Primakoff transformation (HPT)~\cite{holstein_field_1940} for weakly excited two-level emitters would give the same coupled-oscillator model as employed in this manuscript. Thus, the investigation into higher-order expansions of the HPT could be an interesting continuation of the present work.


\begin{acknowledgments}
We would like to thank the anonymous referees whose insightful comments helped to improve our arguments.
The authors acknowledge support from the Swedish Research Council (VR Grant No.~2016-06059).
GJ and AFK acknowledge support from the Knut and Alice Wallenberg Foundation through the Wallenberg Centre for Quantum Technology (WACQT).
AFK acknowledges support from the Swedish Research Council (grant number 2019-03696).
\end{acknowledgments}


%
%
\appendix

\section{Classical coupled oscillator model}
\label{sec:app_COmodel}

In the main text, we compare the results from our quantum model with simulations using a classical coupled-oscillator model which is extensively used in the literature for describing strong coupling. In this appendix, we show the details of how the mapping between the classical model and the quantum model is found.

\subsection{Mapping for an undriven system}

The classical model involves $N+1$ coupled oscillators when $N$ emitters interact with the cavity mode. If all emitters are assumed identical, we may take $\wi = \omega$, $m_i = m$, $k_i = k$ for all emitters with index $i=1,2,..., N$. Letting index $0$ denote the cavity oscillator, the corresponding classical Hamiltonian is written
\begin{align}
    H_{cl} =& \,m_0 \mleft( \frac{p_0^2}{2m_0^2}+ \frac{1}{2}\omega_0^2 x_0^2\mright) \nonumber \\
    &+ \sum^N_{i=1} m \mleft(\frac{p_i^2}{2m^2}+ \frac{1}{2}\omega^2x_i^2 \mright) + \frac{k}{2}\mleft( x_0-x_i \mright)^2 \\
     =& \,m_0 \mleft[ \frac{p_0^2}{2m_0^2}+ \frac{1}{2}\mleft( \omega_0^2+\frac{k}{m_0}\mright) x_0^2\mright] \nonumber \\
    &+ \sum^N_{i=1} m \mleft[ \frac{p_i^2}{2m^2} + \frac{1}{2}\mleft( \omega^2+\frac{k}{m}\mright)x_i^2 \mright] - k x_0 x_i.
    \label{aeq:Hcl}
\end{align}

In the classical model, no rotating-wave approximation (RWA) is applied. To compare with the quantum model, we therefore start with the quantum Rabi Hamiltonian with $N$ identical quantum emitters,
\begin{align}
\hat{H}_{R}&= \hbar\wc \mleft(\ad \a+\frac{1}{2} \mright) \nonumber \\
&+ \sum^N_{i=1}\mleft[ \hbar \we \mleft(\spi \smi-\frac{1}{2} \mright) + \hbar g \mleft(\ad+\a \mright) \mleft(\smi + \spi \mright) \mright] ,
\end{align}
which simplifies to the Tavis--Cummings Hamiltonian when applying the RWA. In a weak-drive approximation, we may replace $\smi \to \bi,\, \spi \to \bid$ with $\bi$ and $\bid$ being the annihilation and creation operators, respectively, of the $i$th quantum oscillator. Then, introducing quantum position and momentum operators $\{\xc, \pc, \xei, \pei\}$, the annihilation operators can be written
\begin{align}
&\a = \sqrt{\frac{\mc \wc}{2\hbar}} \mleft( \xc + i\frac{\pc}{\mc \wc} \mright) \nonumber \\
&\bi = \sqrt{\frac{\me\we}{2\hbar}} \mleft( \xei + i\frac{\pei}{\me \we} \mright). \nonumber
\end{align}
Using these relations, we can write the quantum Hamiltonian in the $\hat{x}\hat{p}$-representation:
\begin{align}
    \hat{H}_{\hat{x}\hat{p}} =&\, \mc\mleft(\frac{\pc^2}{2\mc^2} + \frac{1}{2}\wc^2\xc^2\mright) +\sum^N_{i=1} \me\mleft(\frac{\pei^2}{2\me^2}+ \frac{1}{2}\we^2\xei^2\mright)  \nonumber \\
    &+ 2g\sqrt{\mc\me\wc\we}\xc\xei.
    \label{aeq:Hxp}
\end{align}
Comparing Eqs.~(\ref{aeq:Hcl}) and (\ref{aeq:Hxp}), we can directly identify $m_0 = \mc$, $m = \me$, and 
\begin{align}
 \label{aeq:k}
 & k = - 2g \sqrt{\mc\me\wc \we} ,\\
 \label{aeq:om1} 
 & \omega_0^2 = \wc^2-\frac{k}{\mc} =
 \wc^2+2g\sqrt{\frac{\me\wc\we}{\mc}} ,\\
 \label{aeq:om2}
 & \omega_i^2 = \we^2-\frac{k}{\me} = \we^2+2g\sqrt{\frac{\mc\wc\we}{\me}} .
\end{align}

\subsection{Coherent drive terms}

Finding the classical analog for the strength of an external drive is done in the same manner by comparing the interaction Hamiltonians. Considering a semiclassical model where the quantized cavity mode is driven via dipole interaction with a classical coherent drive $\bar{\mathcal{E}}(t)\cos(\wD t)$, the quantum and classical interaction Hamiltonians are
\begin{align}
     \hat{H}_{d,q} &= -\qc \bar{\mathcal{E}}(t)\cos{(\wD t)}\xc \nonumber\\
    \label{aqe:drive_int}
     & = -\muc \bar{\mathcal{E}}(t)\cos{(\wD t)} \mleft( \ad+\a \mright) ,\\
     H_{d,cl} &= -\qc \bar{\mathcal{E}}(t)\cos{(\wD t)}x_c.
\end{align}
In Eq.~(\ref{aqe:drive_int}), the dipole interaction is rewritten in terms of the transition dipole moment for the cavity oscillator, which in the single-excitation manifold is $\muc=\qc \sqrt{\frac{\hbar}{2\mc\wc}}\hat{e}_k$, where $\hat{e}_k$ is a unit vector along the polarization direction. Then, defining the drive amplitude $\Omd \equiv \frac{\muc \bar{\mathcal{E}}(t)}{\hbar}$, we can write
\begin{align}
     \hat{H}_{d,q}
     & = -\hbar\Omd(t)\cos{(\wD t)} \mleft( \ad+\a \mright) ,\\
     H_{d,cl} &= -\sqrt{2\hbar \mc\wc}\Omd(t)\cos{(\wD t)}x_c.
\end{align}

\section{Quantum theory for the propagating laser beam}
\label{app:Laser}
Inside the laser cavity, the light field is well defined and can readily be quantized. In typical experiments, on the other hand, a beam of light has to travel through free space over distances that are much longer than the characteristic length scales of the studied system. The quantum theory for such a propagating light beam traveling in a straight line in free space can be found in, e.g., Chapter 6 of Ref.~\cite{Loudon_2000} and will be presented briefly below.

Consider a single propagating laser beam under circumstances where transversal effects are irrelevant to the experiment. Then, the quantization geometry can be taken as a finite cross-sectional area $A$ (defined by the experiment) perpendicular to the propagation axis and a quantization axis of infinite length parallel to the propagation axis. This geometry corresponds to a one-dimensional continuous-mode variable that can be taken as the frequency $\omega_k$ with a mode spacing $\Delta \omega = 2 \pi c/L$ that goes to zero as the quantization length $L$ tends to infinity, $L\to \infty$. In this limit, the conversion from sum to integral is
\begin{equation}
    \label{eq:sum_to_int}
    \sum_k \quad \to \quad \frac{1}{\Delta \omega} \int \dd \omega
\end{equation}
and the discrete Kronecker delta is related to a continuous Dirac delta-distribution as
\begin{equation}
    \label{eq:delta}
    \delta_{k,k'} \quad \to \quad \Delta \omega \delta(\omega- \omega').
\end{equation}
It follows that the continuous-mode annihilation and creation operators are related to the discrete operators as
\begin{align}
    &    \a_k \quad \to \quad \sqrt{\Delta \omega}\a(\omega),  \\
    &    \ad_k \quad \to \quad \sqrt{\Delta \omega}\ad(\omega),  
\end{align}
which fulfil the continuous-mode commutation relation,
\begin{equation}
    \comm{\a(\omega)}{\ad(\omega')} = \delta(\omega-\omega').
\end{equation}

Under the assumption of a narrow bandwidth laser, i.e., the excitation bandwidth is much smaller than its central frequency, the lower integration bound can be extended from $0$ to $-\infty$ to cover the entire frequency axis in the integrals above. Thus, the corresponding time-domain operators are obtained as the Fourier transform of $\ad(\omega)$ and $\a(\omega)$:
\begin{align}
    \label{eq:at}
    & \a(t) = \frac{1}{\sqrt{2\pi}}\int^{\infty}_{-\infty} \dd \omega \a(\omega)e^{-i\omega t}, \\
    \label{eq:adt}
    & \ad(t) = \frac{1}{\sqrt{2\pi}}\int^{\infty}_{-\infty} \dd \omega \ad(\omega)e^{i\omega t},
\end{align}
which have the commutation relation,
\begin{equation}
    \comm{\a(t)}{\ad(t')} = \delta(t-t').
\end{equation}
The quantized continuous-mode electromagnetic field operator can then be written as
\begin{equation}
    \hat{E}(z,t) = i \int_0^\infty \dd \omega \sqrt{\frac{\hbar \omega}{4\pi \epsilon_0 c A}}\a(\omega)e^{-i\omega (t-\frac{z}{c})} + {\rm h.c.} 
\end{equation}
where h.c.~denotes Hermitian conjugate. 

The state of the laser field inside the laser cavity can be taken as a coherent state~\cite{Walls_2008}. The output field from the laser will then be a one-dimensional continuous-mode coherent state due to the lack of confinement along the propagation axis. Such a state can be represented using the Fock-space basis kets $\{ \ket{n}\}$ and is created from the vacuum state $\ket{0}$ with the continuous-mode annihilation and creation operators according to
\begin{equation}
    |\{\alpha(t)\}\rangle  = e^{\int \dd \omega \alpha(\omega)\ad(\omega)-\alpha^*(\omega)\a(\omega)}|0\rangle.
\end{equation}
Here $\alpha(\omega)$ is the continuous-mode spectral amplitude. The corresponding time-domain state with wavepacket amplitude $\alpha(t)$ is found via Fourier transform as
\begin{equation}
    |\{ \alpha(t)\} \rangle = e^{\int \dd t [\alpha(t)\ad(t)-\alpha^*(t)\a(t)]}|\text{vac}\rangle.
\end{equation}
The coherent-state mode functions satisfy the normalization condition
\begin{equation}
    \int \dd \omega|\alpha(\omega)|^2 = \int \dd t |\alpha(t)|^2 = \expval{\hat{n}},
\end{equation}
where $\hat{n}$ is the number operator
\begin{equation}
    \label{eq:nhat}
    \hat{n} = \int \dd \omega \ad(\omega)\a(\omega) =  \int \dd t \ad(t)\a(t').
\end{equation}




\subsection{Idealized continuous-wave laser}
In the following, a continuous-wave single-mode laser in a coherent state as described above will be considered. In typical optical experiments, the linewidth of the laser mode is much narrower than the other components of the observed quantum system. Therefore the spectral amplitude can be taken as:
\begin{equation}
    \alpha(\omega) = \sqrt{2\pi}\alpha e^{i\varphi}\delta(\omega-\omega_d).
\end{equation}
Here $\alpha$ is the coherent state amplitude, $\varphi$ the phase, and $\omega_d$ the center frequency. The corresponding wavepacket amplitude $\alpha(t)$ is obtained via the Fourier transform of $\alpha(\omega)$ and is thus a propagating plane wave,
\begin{equation}
    \alpha(t) = \alpha e^{-i\omega_d t + i \varphi}.
\end{equation}
For an ideal stationary beam in a coherent state, 
the mean photon flux, $f(t) = \expval{\ad(t)\a(t)}$, will be time-independent:
\begin{equation}
    f(t)=|\alpha(t)|^2 \equiv \alpha^2.
\end{equation}
As is expected for a stationary beam, with a constant photon flux for all times, the mean photon number $\expval{\hat{n}}$ defined in Eq.~(\ref{eq:nhat}) is infinite, and the spectral amplitude cannot be normalized. These facts make calculations using the quantum representation of the stationary beam problematic.


        

\subsection{Partitioning infinite temporal modes}
The infinite mean photon number is problematic for calculations as it makes the photons in the external drive field ill-defined. A solution to this problem is to define a complete set of discrete, orthonormal basis functions $\{\Phi_i(t)\}$, which partition the continuous-mode coherent laser beam into an infinite tensor product state of discrete-mode coherent states~\cite{Blow_1990}.

If the basis functions $\{\Phi_i(t)\}$ satisfy the orthogonality and completeness relations
\begin{equation}
    \label{aeq:orthogonality}
    \int \dd t\, \Phi_i(t) \Phi_j^*(t) = \delta_{ij}
\end{equation}
and
\begin{equation}
    \label{aeq:completeness}
    \sum_i \Phi_i(t)\Phi_i^*(t') = \delta(t-t'),
\end{equation}
they form a non-continuous basis set with which a discrete set of annihilation operators may be created according to
\begin{equation}
    \hat{c}_i = \int \dd t \Phi_i^*(t)\a(t).
\end{equation}
Equation~(\ref{aeq:completeness}) gives the inverse relation 
\begin{equation}
    \a(t) = \sum_i \Phi_i(t)\ci.
\end{equation}
Naturally, an eigenstate of $\a(t)$ with eigenvalue $\alpha(t)$ is also an eigenstate of $\hat{c}_i$ with eigenvalue
\begin{equation}
    \label{aeq:alpha_i}
    \alpha_i = \int \dd t \,\Phi_i^*(t)\alpha(t).
\end{equation}
It follows that a continuous-mode coherent state can be equivalently expressed as an infinite tensor product of discrete-mode coherent states:
\begin{equation}
    \label{aeq:productstate_prop}
    |\{\alpha(t)\} \rangle = \underset{i}{\Pi}\; e^{\alpha_i \hat{c}_i^\dagger-\alpha^*_i\hat{c}_i}|0\rangle \equiv | \{\alpha_i\}\rangle.
\end{equation}
The result in Eq.~(\ref{aeq:productstate_prop}) is an important property for mode matching the continuous-mode coherent state to a discrete-mode. The freedom in choosing the set of basis functions $\{\Phi_i\}$ is large. This facilitates mode matching of the discrete-mode coherent states $\ket{\{\alpha_i\}}$ with a large variety of mode functions.

One of the simplest examples is the partitioning into rectangular time-bins with duration $T$, which are described by the set of functions $\{\Psi_m(t) \}$ defined as~\cite{van_Enk_2001}
\begin{equation}
    \label{aeq:Psi}
    \Psi_m(t) = 
    \begin{cases}
        \frac{1}{\sqrt{T}} &\text{for} \big|t-\frac{z_0}{c}-mT \big|< \frac{T}{2}\\
        0 & \text{otherwise}
    \end{cases}.
\end{equation}
Above, the label $z_0$ denotes an arbitrarily chosen reference point along the propagation axis. The set of functions in Eq.~(\ref{aeq:Psi}) can be extended to form a complete set that satisfies Eqs.~(\ref{aeq:orthogonality}) and (\ref{aeq:completeness}). The corresponding eigenvalue for each of these discrete-mode coherent states can be obtained from Eq.~(\ref{aeq:alpha_i}) as
\begin{equation}
    \label{aeq:alpha0}
    \alpha_m = \int \dd t \, \Psi_m(t) \alpha(t) \equiv \alpha_0.
\end{equation}
The duration $T$ can be chosen arbitrarily as long as it is much larger than $1/\omega_d$. Thus, the continuous wave laser described by the travelling plane wave $\alpha(t)$ can be expressed as a sequence of $M\to \infty$ copies of the discrete-mode coherent states $|\alpha_0\rangle$ defined by the functions in Eq.~(\ref{aeq:Psi}). 

The benefit of going through all the trouble of reaching this representation is that we now have a well-defined wavepacket amplitude $\alpha_0$ for each partitioned piece of the laser beam.

%
%

\begin{figure}
	\includegraphics[width=.48\textwidth]{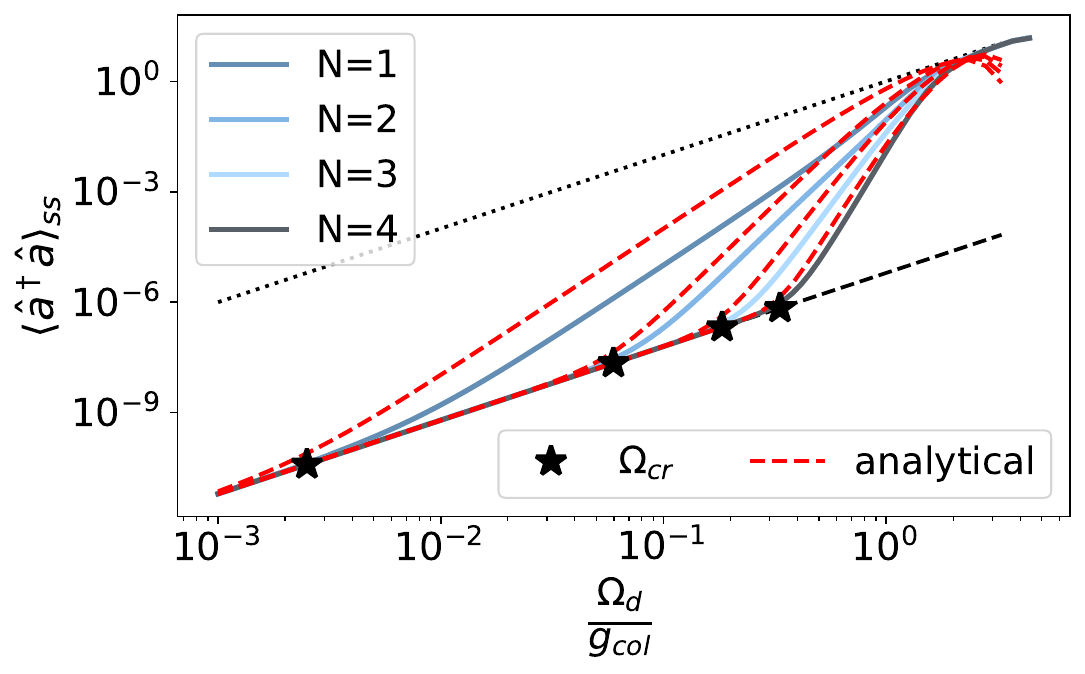}
	\caption{Comparison of the steady-state cavity population calculated with }\eqref{eq:nss_analytical_app}, including the contribution of multiphoton pulses up to order $M=N+5$, and the cavity population obtained with the Lindblad master equation from the main text.
		\label{fig:nss_analytical_app}
\end{figure}

\section{Phenomenological master equation}
\label{app:multiphoton_pulses}
In this appendix, we write down a phenomenological master equation for the probabilities $P_n(t)$ of occupying the $n$th Fock state in the cavity. This master equation provides an analytical approach to gaining an intuition for the $(N+1)$-photon processes associated with the unconventional saturation effect. 
The solution to this master equation was used to obtain the analytical results presented in \figpanel{fig:concept}{b} in the main text.

The setting for the phenomenological master equation is the effective-drive picture described in \secref{sec:EffectiveDrive}. In summary, the cavity population becomes suppressed in the weak-excitation regime due to the coupling to the emitter ensemble. This suppression can be understood by an effective drive, where the external drive and the effective driving from the ensemble interfere destructively and thus cancel the cavity population. However, the ensemble can only cancel photon numbers up to order $N$. Hence, the destructive interference breaks down at order $N+1$. 
The higher-order photon states $(n>N)$ in the external drive can be seen as multiphoton pulses driving the cavity. The effects of such pulses on the cavity dynamics can be studied with a simple phenomenological master equation, which only considers direct cavity absorption and exponential decay. Even though this model neglects all effects from the coupling to the emitters beyond the cancellation of population, it qualitatively captures the unconventional saturation effect, as shown in \figpanel{fig:concept}{b}.

Consider a cavity in the unconventional saturation regime, which is approximately in the ground state because of the destructive interference. Additionally, consider a single pulse with $k$ photons, which is directly absorbed into the cavity due to the intermittent saturation of the destructive interference. 
Neglecting further effects from the coupling to the emitters, the $k$ photons will subsequently leak out of the cavity through single-photon processes with the rate $\gamma_c$ .  In this scenario, the dynamics for the probabilities $P_n$ to occupy the $n$th Fock state can be described by the master equation on vector form
\begin{equation}
	\label{eq:ME_app}
	\partial t \bar{P}(t) = \Gamma \bar{P}(t),
\end{equation}
where $\bar{P}=(P_0, P_1, ... , P_{k})^{\rm T}$ is a column vector with the Fock states $\mleft\{ \ket{n}, \, n\in[0,k] \mright\}$ and
\begin{equation}
\Gamma = \gamma_c \begin{pmatrix}
			0 & 1  & 0  &  	& &\\
			0 & -1 & 2  & 	& &\\
			0 & 0  & -2 &  	& &\\
			   &     &     &  \ddots & \\
			   &	  & 	 & & -(k-1)  & k \\
			   &	  & 	 & & 0	 & 	-k
		\end{pmatrix},
\end{equation}
is a matrix describing the in- and out-flow of probability to occupy each state.
The general solution to \eqref{eq:ME_app} is
\begin{equation}
	\label{eq:ME_app_solution}
	\bar{P}(t) = \exp^{\Gamma t} \bar{P}(0),
\end{equation}
which for $\bar{P}(0)=P_k=1$ gives
\begin{equation}
	\label{eq:Pt}
\bar{P}(t) = \begin{pmatrix}
			\exp^{-k\gamma_c t}\mleft(-1 + \exp^{\gamma_c t}\mright)^k \\
			\binom{k}{1} \exp^{-k\gamma_c t}\mleft(-1 + \exp^{\gamma_c t}\mright)^{(k-1)}\\
			\binom{k}{2} \exp^{-k\gamma_c t}\mleft(-1 + \exp^{\gamma_c t}\mright)^{(k-2)}\\
			\vdots \\
			\binom{k}{k-1} \exp^{-k\gamma_c t}\mleft(-1 + \exp^{\gamma_c t}\mright)^{1} \\
			\exp^{-k\gamma_c t}
		 \end{pmatrix}.
\end{equation}
Equation~(\ref{eq:Pt}) describes the dynamics of the exponential decay of $k$ photons from the $k$th Fock state to the ground state.

\begin{figure*}
	\centering
	    \includegraphics[width = 0.48\textwidth]{DM_one_annotated.pdf}%
	    \includegraphics[width = 0.48\textwidth]{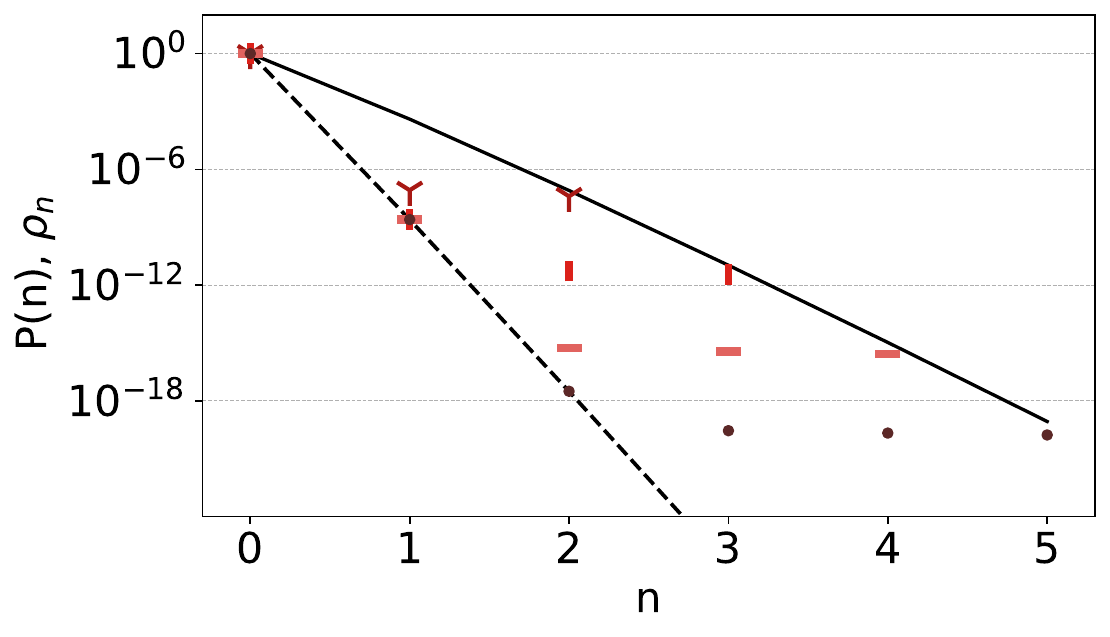}\\
	    \includegraphics[width = 0.48\textwidth]{DM_two_annotated.pdf}%
	    \includegraphics[width = 0.48\textwidth]{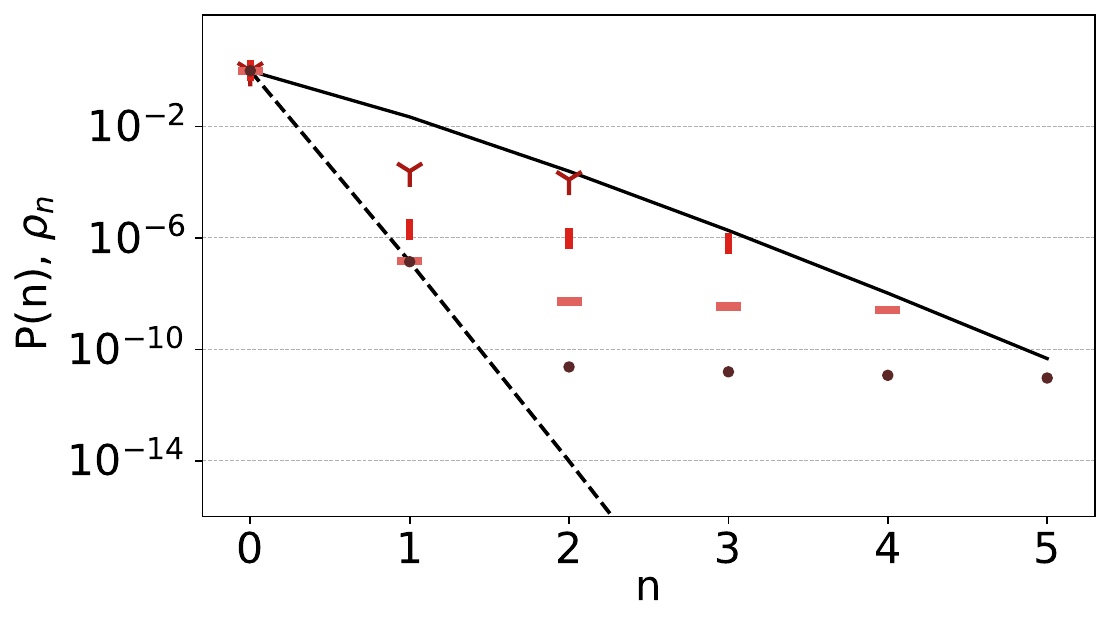}\\	    
	    \includegraphics[width = 0.48\textwidth]{DM_three_annotated.pdf}%
	    \includegraphics[width = 0.48\textwidth]{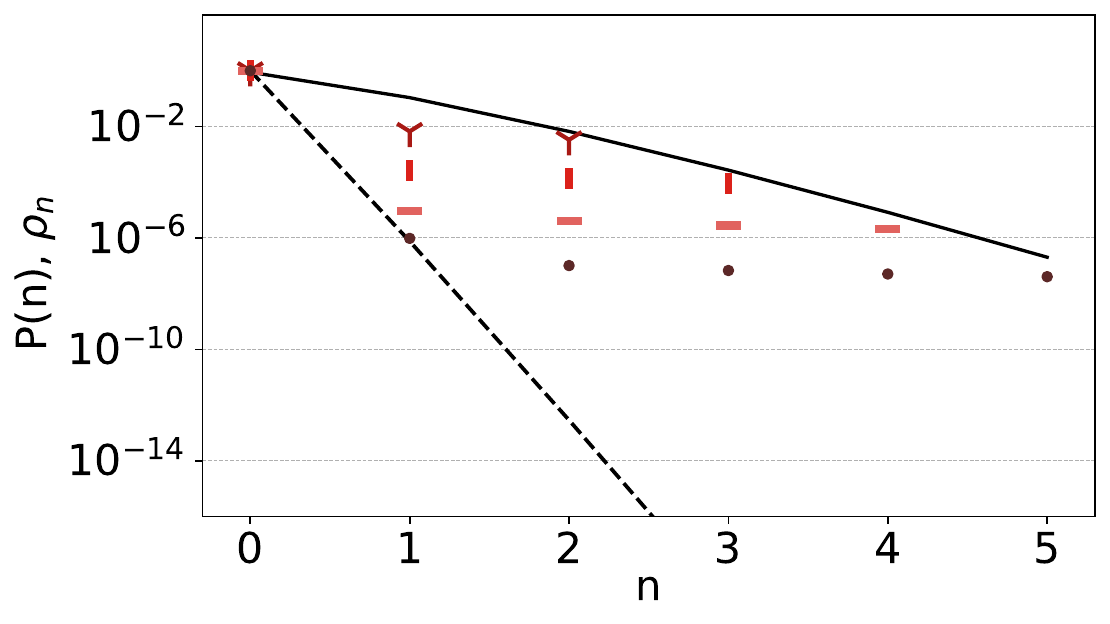} 
    \caption{Comparison of the multiphoton populations $\rho_n$ presented in \figpanels{fig:dm_elements}{b}{c} (left panel) with the analytical prediction based on the phenomenological master equation for $(N+1)$-photon driving (right panel).}
    \label{fig:dm_app}
\end{figure*}

\subsection{Prediction of the total cavity population in the steady-state}
The time-averaged contribution from a stream of independent $k$-photon pulses to the total cavity population in the steady state can be calculated according to
\begin{equation}
\label{eq:nss_kphotons_app}
	\expval{\hat{n}_c}_{ss} = \ncweak+ \frac{1}{T_{ss}}\int_0^{T_{ss}} \dd t \, \frac{T_{ss}}{T} P_{k}^T \sum_{n}^{k} P_n(t).
\end{equation}
The first term is the steady-state cavity population in the weak-drive regime, given by \eqref{eq:nc}. The probability $P_k^T$ of having a $k$-photon pulse during the time $T$ is given by the Poisson distribution for the external drive with amplitude $\alpha_d = \Omega_d T$,
\begin{equation}
\label{eq:PkT}
	P_k^T = \exp^{-\abs{\Omd T}^2}\frac{\abs{\Omd T}^{2k}}{k!}.
\end{equation}
The factor $T_{ss}/T$ gives the number of time bins with duration $T$ in the time $T_{ss}$, over which we average. Taken together, the term $(T_{ss}/T) P_k^T$ gives the fraction of $k$-photon pulses that contributes to the time average. Using the explicit forms for the time-dependent probabilities $P_n(t)$ given in \eqref{eq:Pt}, the sum in \eqref{eq:nss_kphotons_app} can be evaluated as $\sum_{n=0}^k nP_n(t) = k\exp^{-\gamma_c t}$. Additionally, we can assume that the time it takes to reach the steady state is much longer than the time scale for the decay dynamics, $T_{ss}\gg 1/\gamma_c$. Thus, we can take the limit $T_{ss} \to \infty$, which yields the expression
\begin{equation}
\label{eq:nss_analytical_app}
	\expval{\hat{n}_c}_{ss} = \ncweak+  \frac{kP_k^T}{T \gamma_c}.
\end{equation}
When $N$ emitters couple to the cavity, the destructive interference between the emitters and the external drive breaks down at order $N+1$. Hence, the lowest-order photon pulses contributing to the cavity population in the steady state will be $N+1$. Taking $k = N+1$ in \eqref{eq:nss_analytical_app} gives the expression for the cavity population given in \eqref{eq:nss_analytical} in the main text.

Equation\,(\ref{eq:nss_analytical_app}) can easily be generalized to include the contribution from independent $k$-photon pulses up to order $M$,
\begin{equation}
\label{eq:nss_analytical_app_M}
	\expval{\hat{n}_c}_{ss} = \ncweak+  \sum_{k=N+1}^M \frac{kP_k^T}{T \gamma_c}.
\end{equation}
The cavity populations for $N=1-4$, calculated with \eqref{eq:nss_analytical_app_M} and  $M=N+5$, are plotted as dashed red curves in \figref{fig:nss_analytical_app}. Solid blue curves show the results calculated with the Lindblad master equation from the main text. A comparison of \figref{fig:nss_analytical_app} with \figpanel{fig:concept}{b} in the main text shows that including a few more orders of multiphoton absorption in the simple model described in this section qualitatively captures the unconventional saturation effect for drive strengths close to the traditional saturation point $\Omd \sim \gcol$. The extension does \textit{not}, however, change the behaviour in the intermediate drive regime $(\Omd>\gcol)$.

\subsection{Prediction of the cavity populations in the steady state}
Using the analytical solutions in \eqref{eq:Pt}, we can also find an expression for the probability of occupying the $n$th Fock state in the steady state. Including only the contribution from $(N+1)$-photon pulses, the expression is
\begin{equation}
\label{eq:nss_kphotons}
	\expval{P_n}_{ss} = \expval{P_n}_{\rm weak}+ \frac{1}{T_{ss}}\int_0^{T_{ss}} \dd t \, \frac{T_{ss}}{T} P_{N+1}^T P_n(t).
\end{equation}
Performing the integral with $P_n(t)$ given by the solution to the master equation given in \eqref{eq:Pt}, and taking the limit $T_{ss}\to \infty$, gives
\begin{equation}
\label{eq:Pn}
	\expval{P_n}_{ss} = \expval{P_n}_{\rm weak}+ \frac{ P_{N+1}^T}{nT\gamma_c}.
\end{equation}
The probability $P_{N+1}^T$ is given by \eqref{eq:PkT} with $k=N+1$ and $\expval{P_n}_{\rm weak}$ is given by the Poisson distribution for a coherent population with amplitude $\abs{\alpha_{\rm weak}}^2=\ncweak$,
\begin{equation}
\label{eq:Pn_weak}
	 \expval{P_n}_{\rm weak}  = \exp^{-\abs{\alpha_{\rm weak}}^2}\frac{\abs{\alpha_{\rm weak}}^{2n}}{n!}.
\end{equation}

The second term in \eqref{eq:Pn} is proportional to $\Omega_d^{2(N+1)}$. Thus, this term explains the cross-over from the coupled distribution to the uncoupled distribution [$P_{\alpha_c}(n)$ and $P_{\alpha_c^o}(n)$ in \figref{fig:dm_elements}, respectively] at $n=N+1$. In the weak-drive regime, before the  system has entered the unconventional saturation regime, the first term $\expval{P_n}_{\rm weak}$ will be much larger than the contribution from the $(N+1)$-photon pulses for $n< N+1$. Therefore, the $\expval{P_n}_{\rm ss}$ will still have a Poisson-like distribution, but will deviate from the coupled distribution because of the contribution from the $(N+1)$-photon pulses given by the second term in \eqref{eq:Pn}. On the other hand, when the system is in the unconventional saturation regime, the probability $P_{N+1}^T$ of having $(N+1)$ photons in the drive has become large enough for the second term in \eqref{eq:Pn} to dominate. Hence giving $\expval{P_1}_{\rm ss}\approx\expval{P_2}_{\rm ss} ... \approx \expval{P_{N+1}}_{\rm ss}$. Thus, \eqref{eq:Pn} explains the observation of the relationship $(N+1)\rho_{N+1} \approx N\rho_{N} \approx ... \approx \rho_1$ between the populations in the unconventional saturation regime.

In \figref{fig:dm_app}, we compare the $\expval{P_n}_{ss}$ given in \eqref{eq:Pn} (right panel) with the cavity populations $\rho_n \equiv \expval{\rhoh_c}{n}$ from \figref{fig:dm_elements}[(b),(c),(d)] (left panel). This comparison shows that the simple phenomenological master equation described in this appendix gives a good analytical intuition to the cavity response to external driving observed in the main text. The analytical results also confirm the expectations on the populations $\rho_n$ discussed in \secref{sec:suppression}.

%
%

\begin{figure}
    \centering
    \includegraphics[width=.48\textwidth]{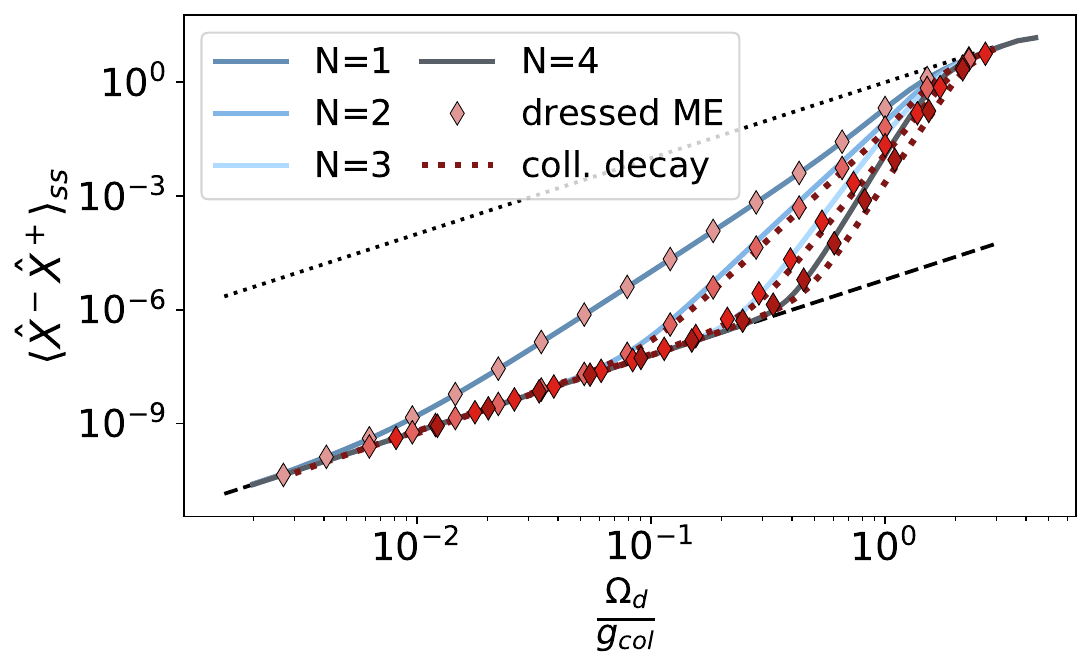} %
    \caption{Visualization of the effects of counter-rotating terms (diamonds and red dotted curves) and collective emitter dissipation (red dotted curves) on the unconventional saturation effect. Solid curves reproduce the results of \figpanel{fig:ss(Om)}{a} from the main text.}
    \label{fig:ME_comparison}
\end{figure}

\section{Master equations including counter-rotating terms and collective emitter decay}
\label{app:RWA}
Section \ref{sec:limitations} of the main text discussed the validity of the master equation employed in this work. In \figref{fig:ME_comparison}, we show how the mentioned alternative master equations affect the cavity response to external driving with the current set of parameters. 
To better facilitate the comparison, we again write down the master equation in \eqref{eq:ME}:
\begin{equation}
	\label{eq:ME_main}
        \dot{\rhoh} = -\frac{i}{\hbar} \mleft[\H_{TC},\rhoh\mright] + \gamc \mathcal{D}_{\a}\mleft[\rhoh\mright] + \sum_{i=1}^N \game \mathcal{D}_{\smi}\mleft[\rhoh\mright].
\end{equation}

Which master equation is correct when the RWA is retracted strongly depends on the set of parameters. Because of the secular approximation, the commonly used
master equation in the USC regime~\cite{beaudoin_dissipation_2011} is suitable for the good-cavity regime ($g_{col} \gg \gamma_c$), where the transitions of the system can be assumed non-overlapping. The secular approximation is, however, not suitable for the lossy-cavity regime studied in this manuscript. Because of the cavity dissipation, the overlap of the transitions in this regime cannot be neglected. Instead, we have performed a similar derivation as in Ref.~\cite{settineri_dissipation_2018,rau_cavity_2004} to obtain the following master equation: 
\begin{align}
	\label{eq:ME_corr}
    \dot{ \hat{\rho}} =
	 &  -\frac{i}{\hbar} \comm{\hat{H}_{TC} + \sum_{i=1}^N \hbar g\left( \hat{a}^\dagger \hat{\sigma}_{+i} + \hat{a} \hat{\sigma}_{-i} \right) }{\hat{\rho}} \nonumber \\
	& + \gamma_c \mathcal{D}_{\hat{X}^+_c}[\hat{\rho}] + \sum_{i=1}^N \gamma_e \mathcal{D}_{\hat{X}^+_{ei}}[\hat{\rho}].
\end{align}
Here, the counter-rotating terms $\hat{a}^\dagger \hat{\sigma}_{+i}$ and $\hat{a} \hat{\sigma}_{-i}$ are included in the Hamiltonian, and $\mathcal{D}_{\hat{o}}[\cdot] = \hat{o} \cdot \hat{o}^\dagger -1/2 \acomm{\hat{o}^\dagger \hat{o}}{\cdot}$. The positive-frequency operators $\hat{X}^+_c$ and $\hat{X}^+_{ei}$ are defined as
\begin{align}
       \label{eq:cavity_diss}
	&	\hat{X}^+_c \equiv \sum_{j,k>j} \matrixel{j}{\hat{a}^\dagger + \hat{a}}{k} \ketbra{j}{k} \\
	\label{eq:emitter_diss}
	&     \hat{X}^+_{ei}\equiv \sum_{j,k>j} \matrixel{j}{\hat{\sigma}_{+i} + \hat{\sigma}_{-i}}{k} \ketbra{j}{k},
\end{align}
where $\ket{j}$ and $\ket{k}$ are eigenstates to the undriven Tavis--Cummings Hamiltonian before making the RWA.
This master equation brings the system to the correct dressed ground state, including counter-rotating terms, and is suitable for systems with overlapping transitions.

The derivation of a collective dissipator, including counter-rotating terms, is analogous to the derivation of \eqref{eq:emitter_diss}. The result can be directly written down by replacing $\hat{\sigma}_{+i}$ and $\hat{\sigma}_{-i}$ with the collective pseudo-spin operators $S_+\equiv \sum_{i=1}^N \hat{\sigma}_{+i}$ and $S_-\equiv \sum_{i=1}^N \hat{\sigma}_{-i}$. The corresponding collective dissipation rate is $\gamma_{col}=\gamma_e/N$.

In \figref{fig:ME_comparison}, we compare the cavity response to external driving obtained with the different master equations presented above. Solid, blue, and grey curves show the cavity response to external driving for $N = 1-4$ quantum emitters obtained with \eqref{eq:ME_main}, which is suitable for the Tavis--Cummings model and individual emitter decay as presented in the main text. Red diamonds show the cavity response calculated with counter-rotating terms in the cavity-emitter interaction and individual decay of the emitters. Lastly, dotted red curves show the cavity response calculated with counter-rotating terms and collective decay of the emitters. As can be seen, the presented changes to the master equation do \textit{not} affect the unconventional saturation effect qualitatively. Thus, the RWA captures all the essential physics we demonstrate in our manuscript.

Note that the quantity plotted on the y-axis is $\expval{\hat{X}^- \hat{X}^+}_{ss}$ and not directly $\expval{\hat{a}^\dagger \hat{a}}_{ss}$. The operators $\hat{X}^+=\sum_{j,k>j}\matrixel{j}{\hat{a}^\dagger + \hat{a}}{k}\ketbra{j}{k}$ and $\hat{X}^-=(\hat{X}^+)^\dagger$ correspond to the positive and negative frequency component of the photon-like field inside the cavity, respectively.
With a correct treatment of input-output theory, the field emitted from the cavity is proportional to $\hat{X}^+$~\cite{ridolfo_photon_2012,Kockum2019}. Thus, $\hat{X}^+$ is the relevant field operator for experiments. When the RWA is applied, $\hat{X}^+=\hat{a}$ when the cavity and emitters are on resonance. On the other hand, when the RWA is not applied, $\hat{X}^+=\hat{X}^+_c$, with $\hat{X}^+_c$ given in~\eqref{eq:cavity_diss}. Thus, the notation $\expval{\hat{X}^- \hat{X}^+}_{ss}$ allows us to compare the results obtained with the different master equations. Following this note,~\figref{fig:ME_comparison} further tells us that the unconventional saturation effect appears in the excitation of the photon-like field (counted with the operator $\hat{X}^- \hat{X}^+$) that leaks out of the cavity and not directly in the population of the cavity Fock states (counted with the operator $\hat{a}^\dagger \hat{a}$). Therefore, there is no distinction between applying or not applying the RWA for experiments.

To obtain the cavity response with counter-rotating terms in the Hamiltonian (diamonds and dotted curves in \figref{fig:ME_comparison}), we have slightly changed the drive frequency $\omega_d$. Including the counter-rotating terms in the Hamiltonian, we find a minor shift of the transparency dip observed in \figpanel{fig:spectrum}{b}. This shift slightly changes the drive frequency at which the effect is most pronounced. With the current set of parameters, we find that this shift is less than \unit[0.5]{\%} for $N=1$ and less than \unit[0.3]{\%} for $N=4$. Since this shift is so small, not correcting the drive frequency will still show the effect and we do not find any qualitative changes in the unconventional saturation effect.

\section{Cooperativity}
\label{app:add_sim}

\label{app:coop}

\begin{figure*}
    \centering
    \includegraphics[width = \textwidth]{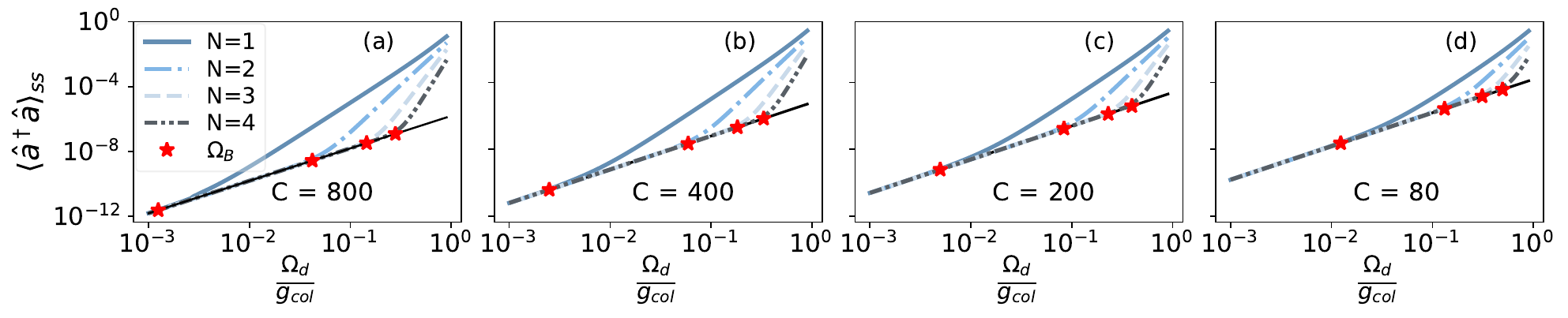} \\
    \includegraphics[width = \textwidth]{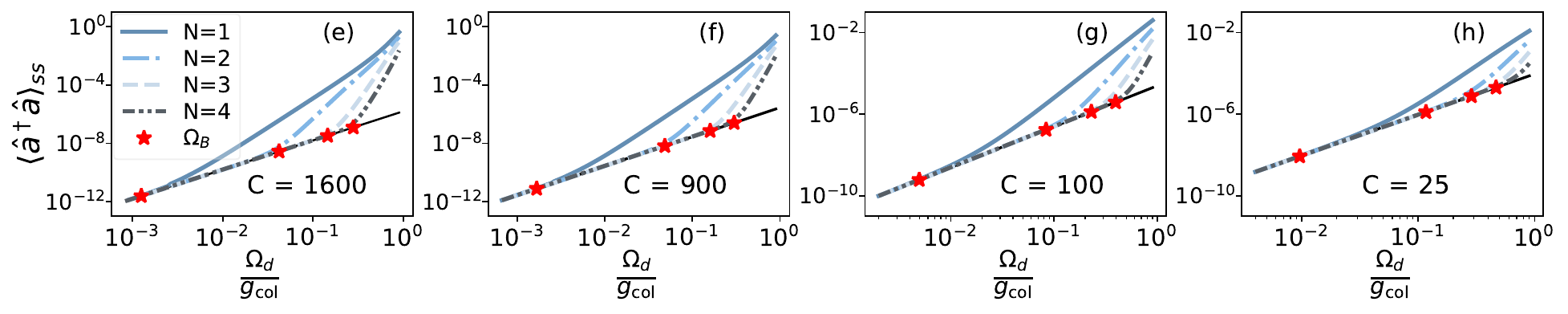}
    \caption{The critical drive panels from the main text with the corresponding cooperativities $C$. In panel \textbf{(a\,-\,d)}, $\gcol$ and $\gamc$ have been held fixed while $\game$ has been varied from $0.5\,$-$5\%$ of $\gamc$ going left to right. In panel \textbf{(e\,-\,h)}, $\gamc$ and $\game$ have been held fixed while $\gcol$ has been varied from $2\,$-$0.25\,\gamc$ going left to right. } 
    \label{fig:Coop_panels}
\end{figure*}

The cooperativity is defined as
\begin{equation}
    \label{aeq:coop}
    C \equiv \frac{4\gcol^2}{\gamc \game}.   
\end{equation}
In \figref{fig:Coop_panels}, the panels with break-point predictions from \figref{fig:OmB} in the main text are shown together with the corresponding cooperativities. As can be seen, the observed saturation effect and our analytical expression for the critical drive are both robust to a wide range of cooperativities, $25\leq C \leq 1600$. It is also evident from \figref{fig:Coop_panels}, that a large cooperativity facilitates the observation of the saturation effect as it pushes the emergence of the nonlinear effect to lower drive strengths.

\begin{figure*}
    \centering
    \includegraphics[width=.48\textwidth]{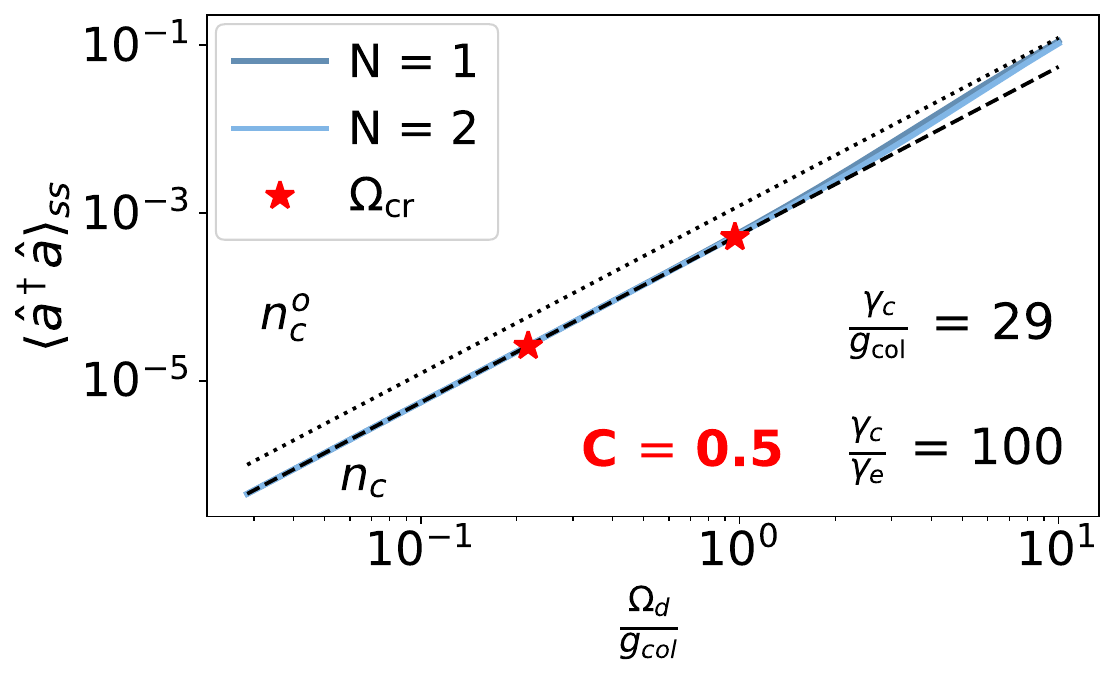} %
    \includegraphics[width=.48\textwidth]{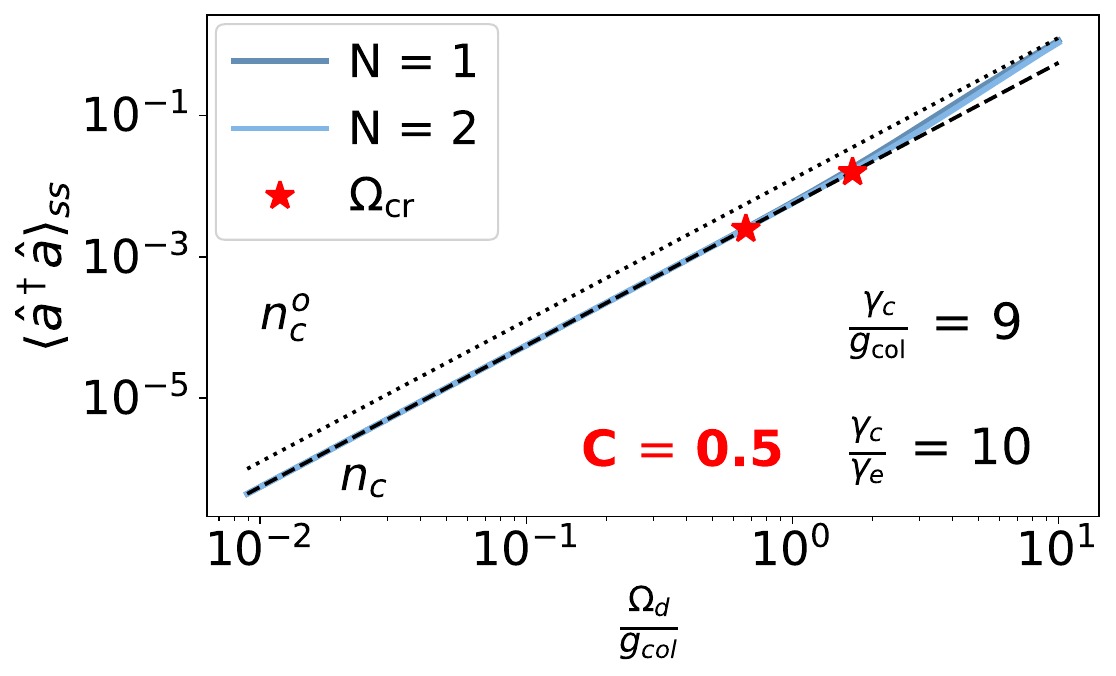} \\
    \includegraphics[width=.48\textwidth]{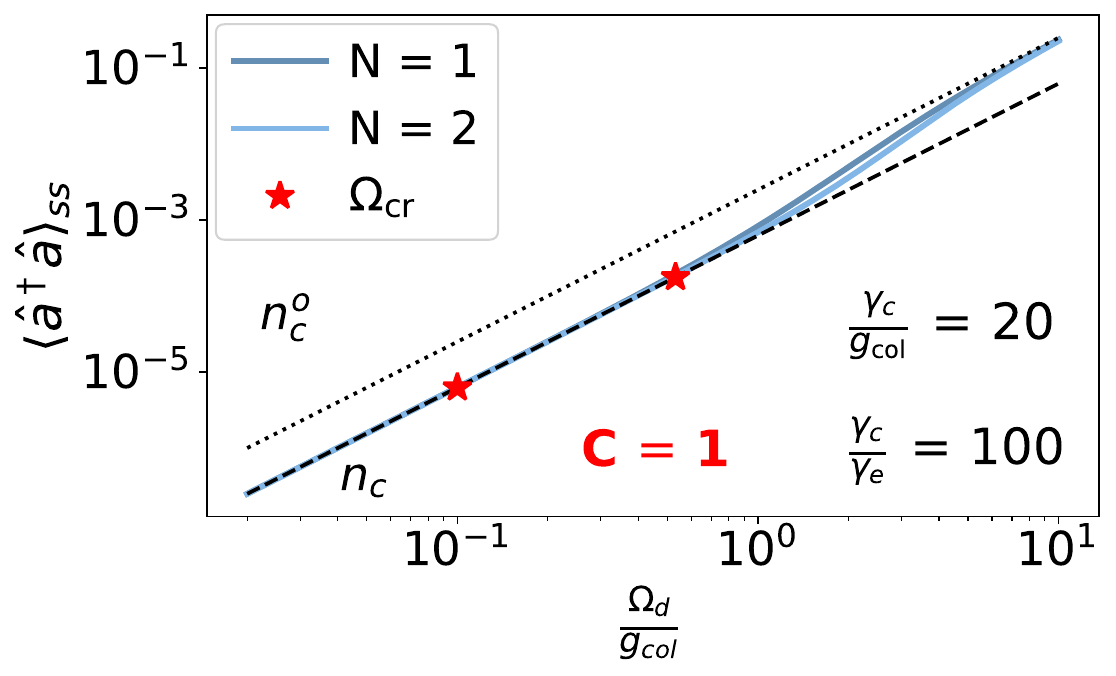} %
    \includegraphics[width=.48\textwidth]{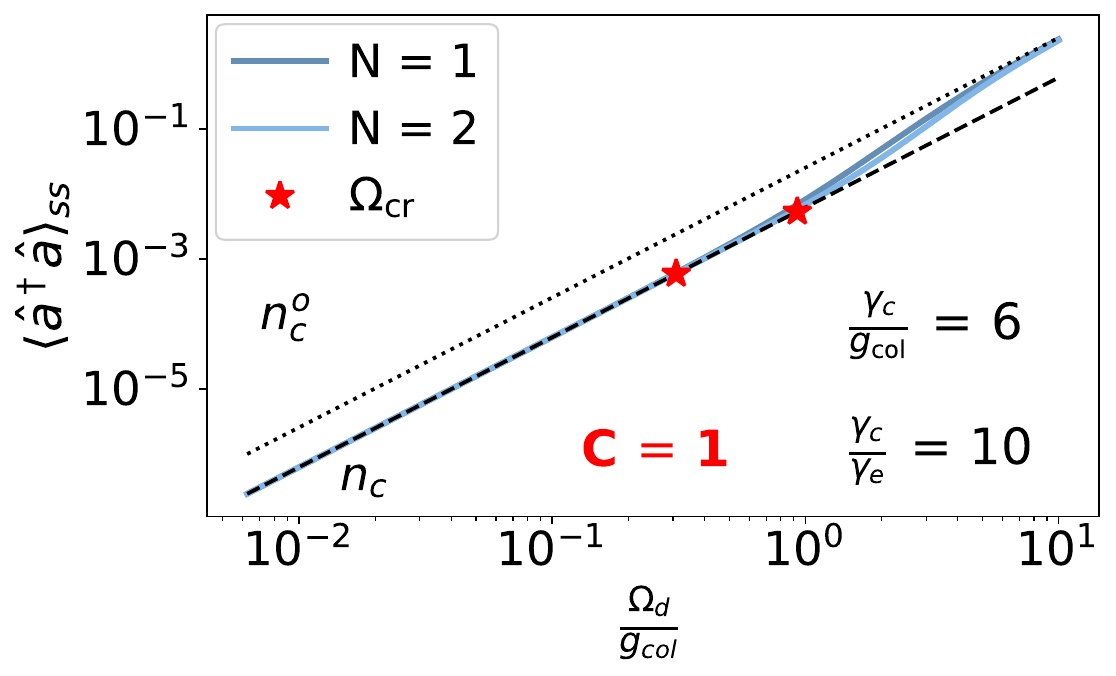} \\
    \includegraphics[width=.48\textwidth]{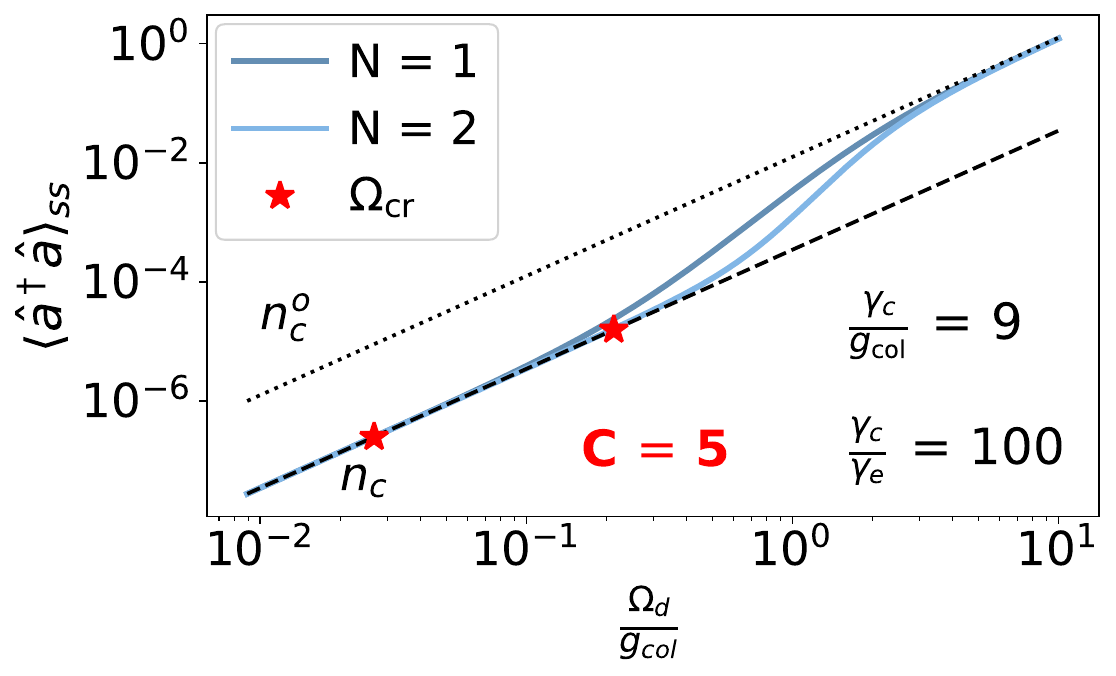} %
    \includegraphics[width=.48\textwidth]{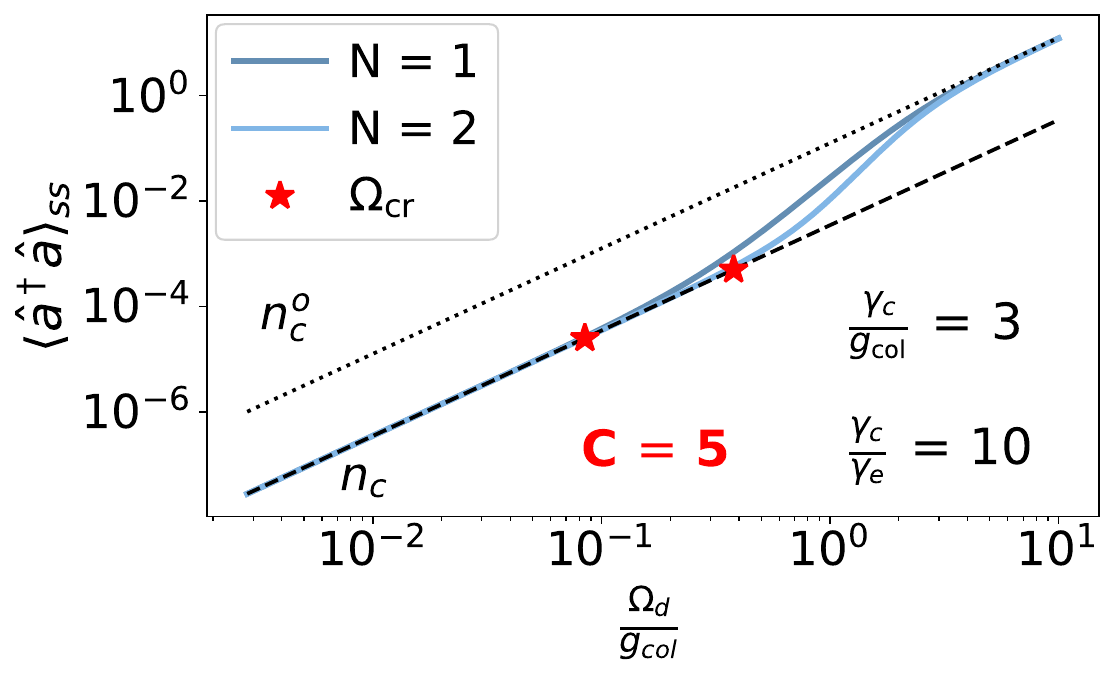} \\
    \includegraphics[width=.48\textwidth]{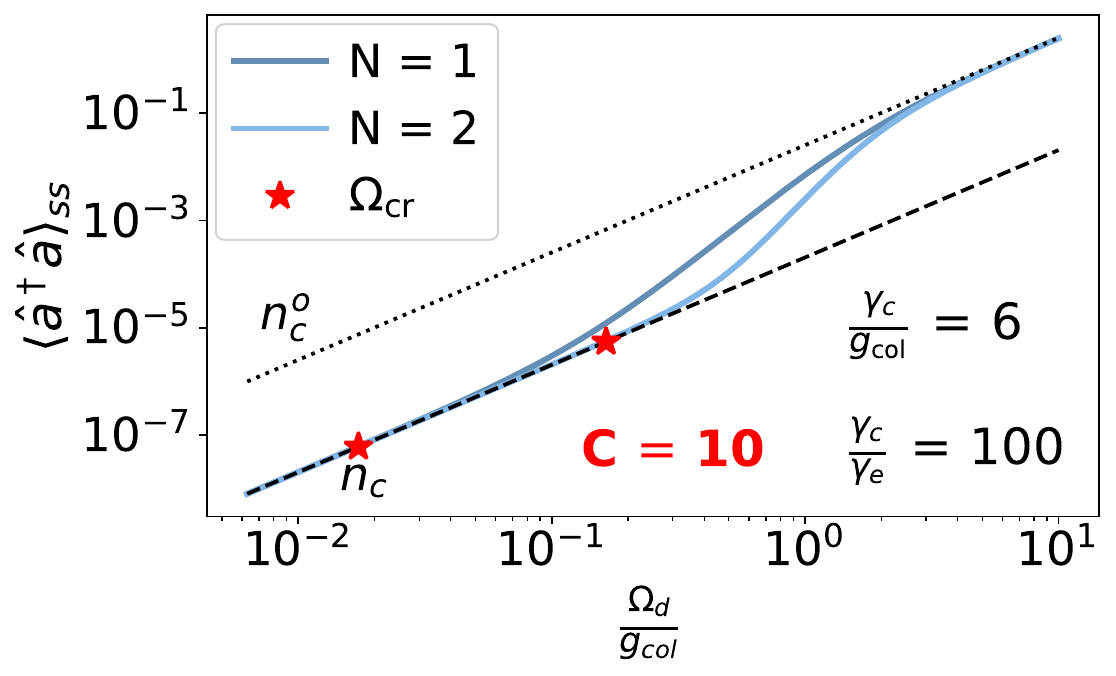} %
    \includegraphics[width=.48\textwidth]{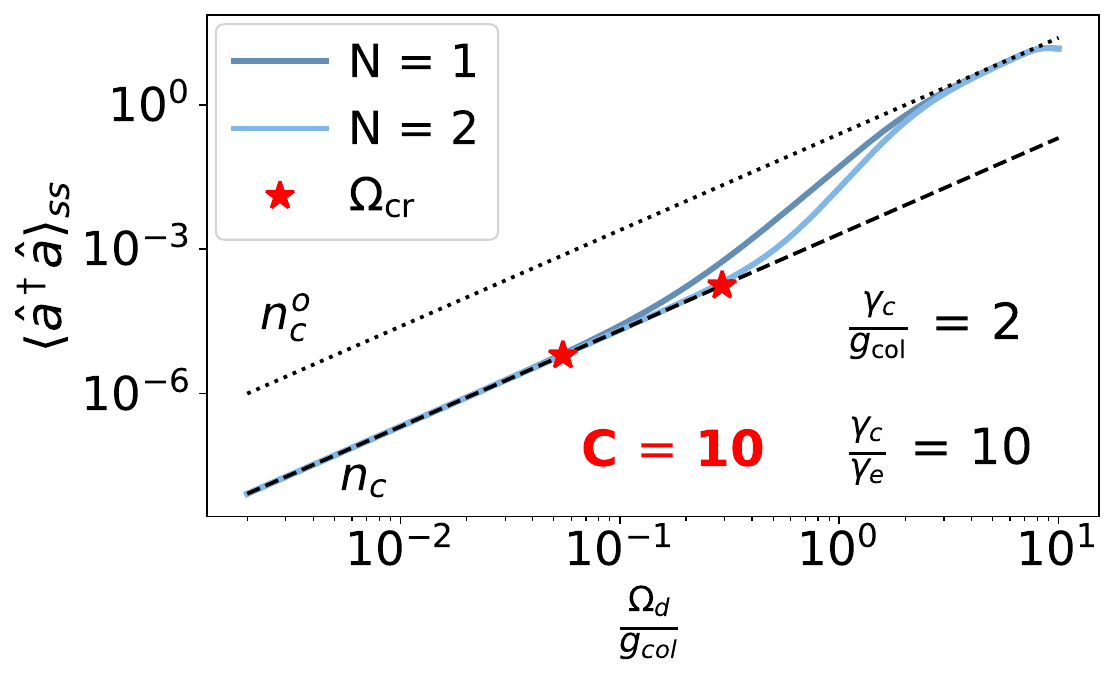} \\
    \caption{Steady-state cavity populations as a function of drive strength for cavity loss rate $\gamc\approx 0.03\,\wc$ and varying emitter loss rate and collective coupling.}
    \label{fig:coop_set1}
\end{figure*}

\begin{figure*}
    \centering
    \includegraphics[width=.48\textwidth]{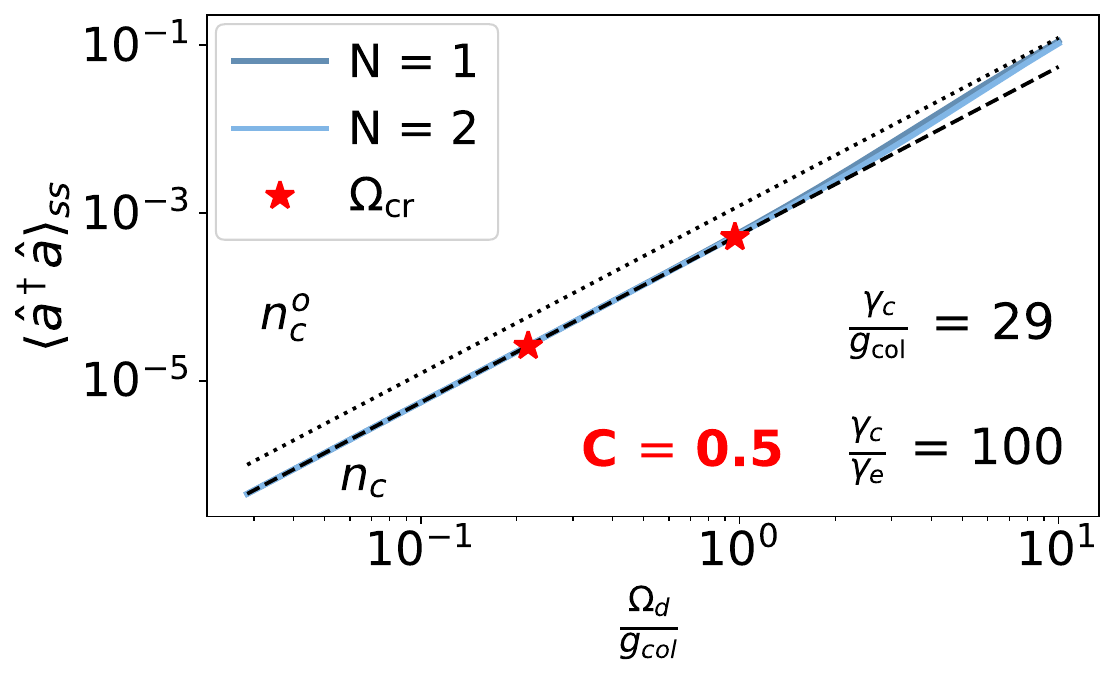} %
    \includegraphics[width=.48\textwidth]{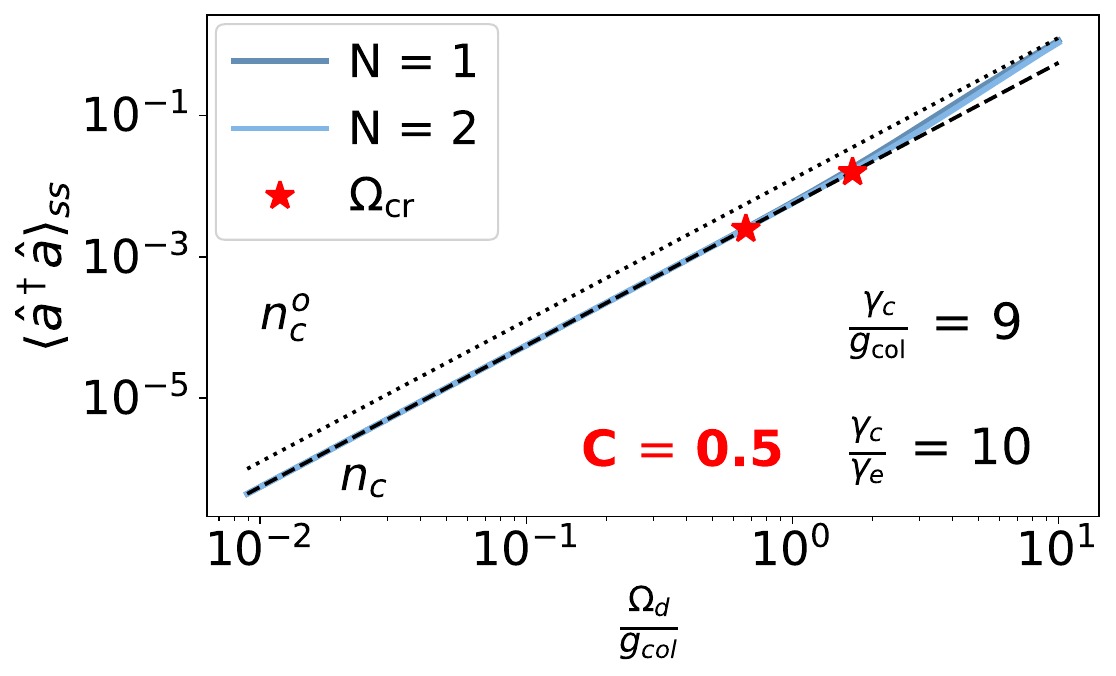} \\
    \includegraphics[width=.48\textwidth]{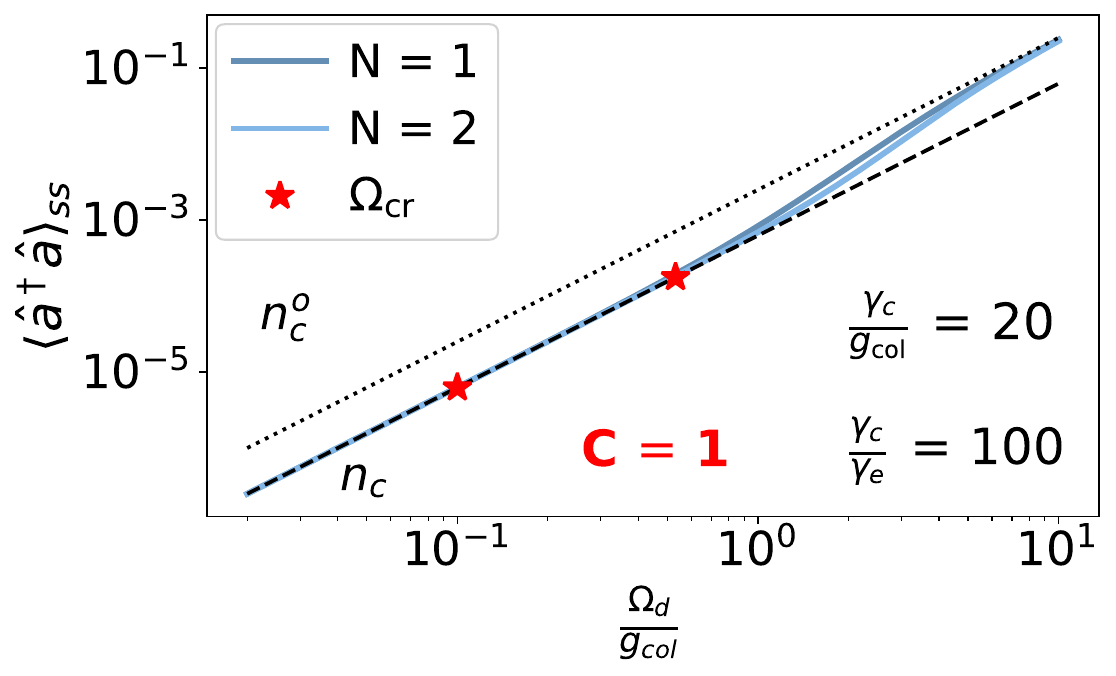} %
    \includegraphics[width=.48\textwidth]{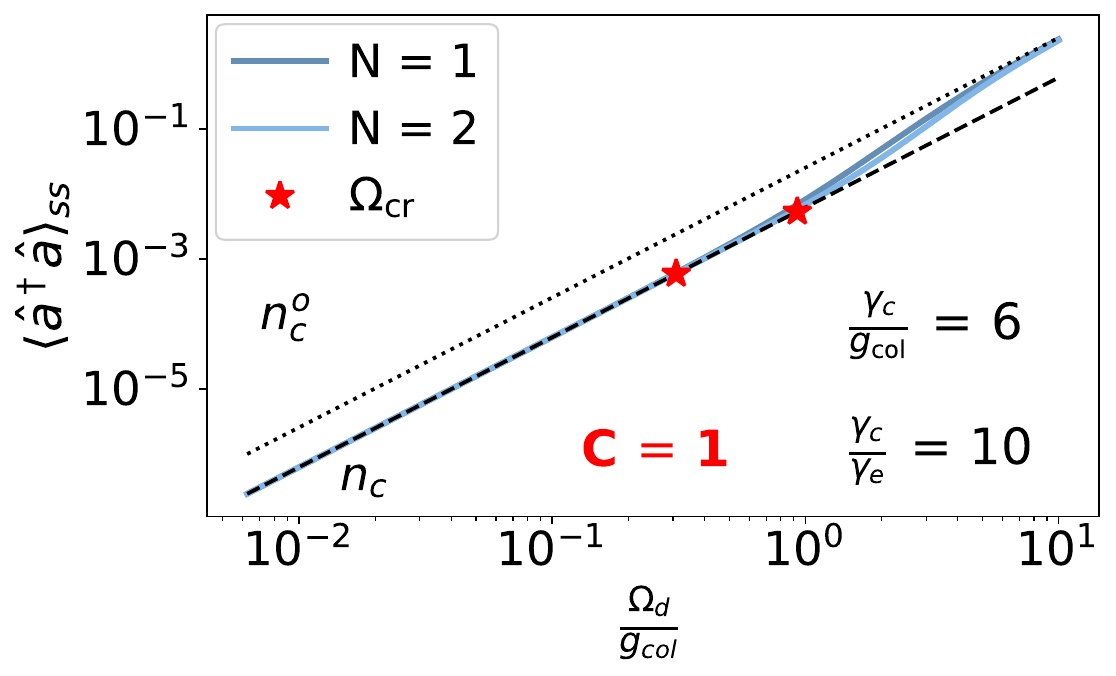} \\
    \includegraphics[width=.48\textwidth]{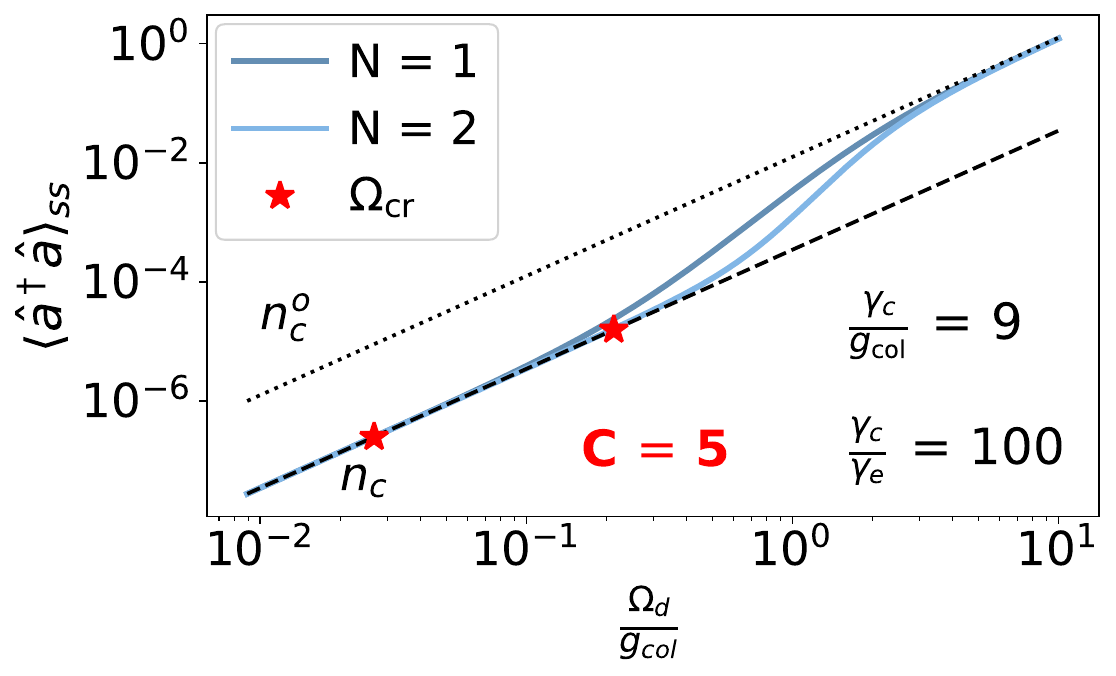} %
    \includegraphics[width=.48\textwidth]{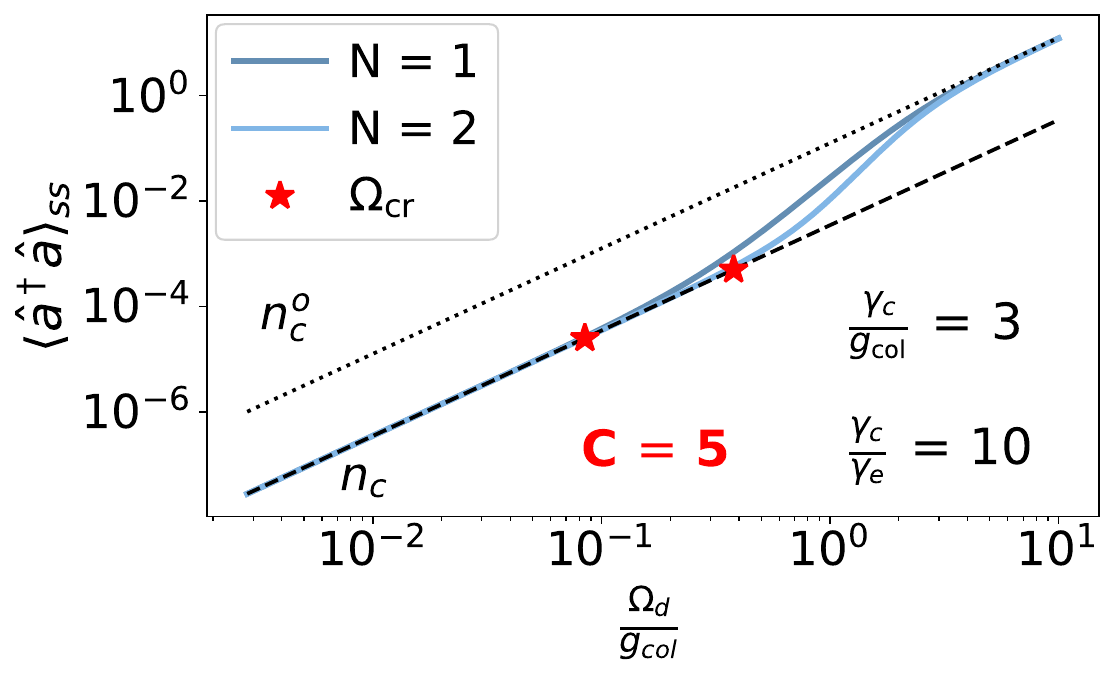} \\
    \includegraphics[width=.48\textwidth]{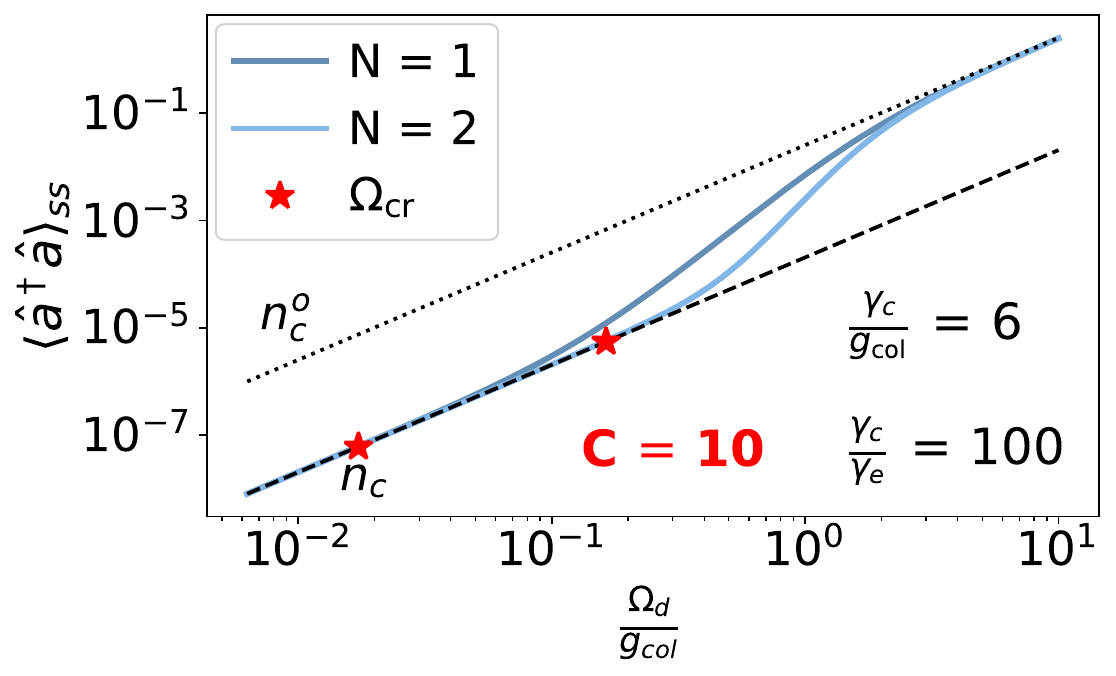} %
    \includegraphics[width=.48\textwidth]{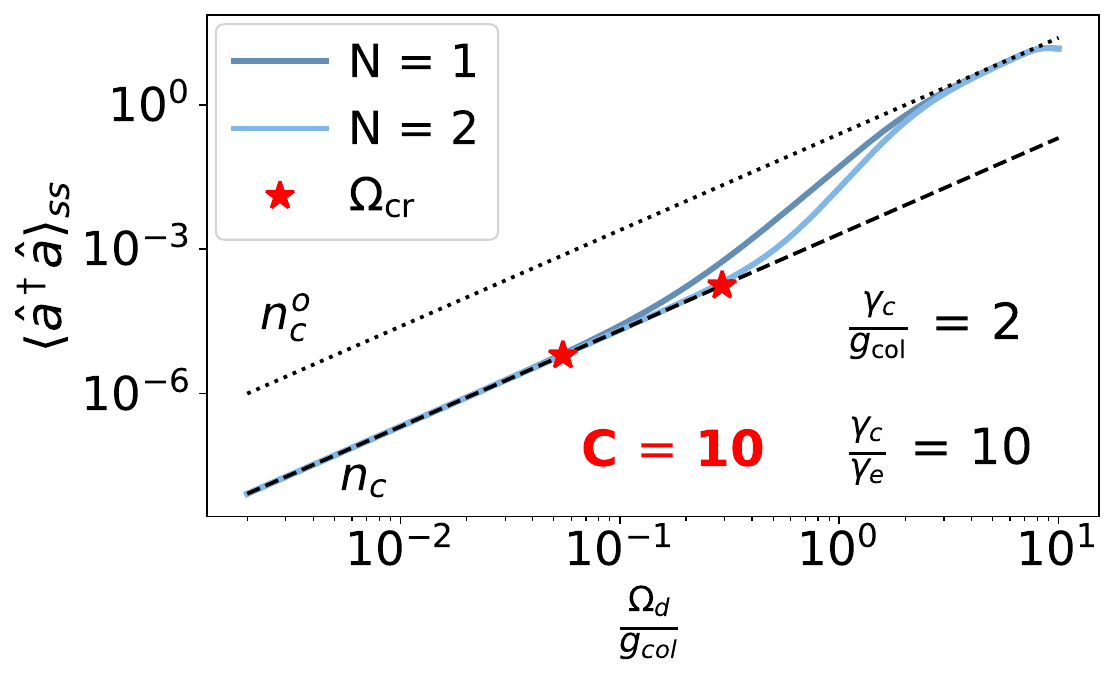} \\
    \caption{Steady-state cavity populations as a function of drive strength for cavity loss rate $\gamc\approx 0.17\,\wc$ and varying emitter loss rate and collective coupling.}
    \label{fig:coop_set2}
\end{figure*}

To find out how robust the observed effect is at low cooperativity, we performed a few simulations with one and two emitters in the cavity for different $C$. The results from these investigations are presented below and are structured as follows. First, a set of plots with a semi-lossy cavity,  i.e., $\gamc\approx 0.03\,\wc$, is presented in \figref{fig:coop_set1}. This cavity loss rate is the same as was used to produce the plots in \figref{fig:Coop_panels}. After that, a second set of plots, where $\gamc\approx 0.17\,\wc$, is presented in \figref{fig:coop_set2}. This loss rate corresponds, e.g., to a localised surface plasmon mode in a metal nanoparticle. 
Moreover, both \figref{fig:coop_set1} and \ref{fig:coop_set2} have two columns. The first column shows the response with low-loss emitters, $\game = 0.01\,\gamc$, and the second column shows intermediate-loss emitters with $\game = 0.1\,\gamc$. According to \eqref{aeq:coop}, a more lossy emitter ensemble can be compensated to have the same cooperativity as a less lossy one ($\gamc$ fixed) by increasing the cavity-emitter interaction strength. Therefore, the two columns could also be regarded as corresponding to very weak coupling in the left column, and weak coupling in the right column.


%
%
\FloatBarrier

\bibliography{references}

\end{document}